\documentclass[12pt]{article}

\usepackage{amsmath,amssymb}

\newcommand{\ep}{\varepsilon}
\DeclareMathOperator{\trace}{trace}

\newbox\shell
\newcommand{\dia}[2]{\setbox\shell=\hbox{\begin{picture}(180,220)(-90,-110)#1
\put(-90,-110){\makebox(180,220)[b]{\large #2}}\end{picture}}\dimen0=\ht
\shell\multiply\dimen0by7\divide\dimen0by16\raise-\dimen0\box\shell\hfill}
\newcommand{\vtx}{\circle*{10}}

\begin{document}

{\large \begin{center}
{\bf Feynman integrals, L-series and Kloosterman moments}\\[10pt]
David Broadhurst, Open University, UK\\[10pt]
20 February 2016
\end{center}

This work lies at an intersection of three subjects: quantum field theory,
algebraic geometry and number theory, in a
situation where dialogue between practitioners
has revealed rich structure. It contains a theorem and 7 conjectures,
tested deeply by 3 optimized algorithms, on
relations between Feynman integrals and L-series defined by products, over the primes,
of data determined by moments of Kloosterman sums in finite fields.
There is an extended introduction, for readers who
may not be familiar with all three of these subjects. Notable new results
include conjectural evaluations of non-critical L-series of modular forms 
of weights 3, 4 and 6, by determinants of Feynman integrals, an evaluation
for the weight 5 problem, at a critical integer, and  formulas for determinants
of arbitrary size, tested up to 30 loops. It is shown that the functional
equation for Kloosterman moments determines much but not
all of the structure of the L-series.   In particular, for problems with 
odd numbers of Bessel functions, it misses a crucial feature captured 
in this work by novel and intensively tested conjectures. For the 9-Bessel
problem, these lead to an astounding compression of data at the primes.}

\newpage

{\large \section{Introduction}%1

As indicated in the abstract, this article concerns a fruitful intersection of three fascinating subjects: 
quantum field theory~\cite{BLinz}, 
algebraic geometry~\cite{BKV} and 
number theory~\cite{GZ}, with a focus on Feynman integrals
that evaluate to L-series defined by moments of Kloosterman sums~\cite{R}. 

Feynman diagrams with internal
edges associated to massive particles invariably define {\em integrals of products of Bessel
functions.} Simple cases were studied in~\cite{BBBG}. 
{\em Kloosterman moments} are mere integers, resulting from
finite sums in finite fields. Bridges between these seemingly
different constructs are provided by {\em L-series}. These are defined by products over all the primes,
in the manner that Euler obtained zeta values by considering
\begin{equation} \zeta(s)=\prod_p\frac{1}{1-p^{-s}}=\sum_{n>0}\frac{1}{n^s}.
\label{zeta}\end{equation}
Key data comes from {\em functional equations}~\cite{FW3} for Kloosterman moments, 
at powers of primes. These yield functional equations for L-series, similar to that used
by Riemann to study the analytic continuation of $\zeta(s)$ inside a critical strip and most
spectacularly on a critical line, with the real part of $s$ set to $\frac12$.
Our strips will be wider, containing {\em critical integers}, on the real line. At these, {\em mirabile dictu},
L-series evaluate to Feynman integrals~\cite{BLinz,BS}.

The remainder of this introduction provides further orientation, for those who wish it.
A reader who finds fun in formulas may fast-forward,
taking note that each of the 24 signs $\stackrel{?}{=}$ in Section~7 indicates an empirical result that is, 
to my knowledge, unproven, yet has been tested numerically to a precision of at least 500 decimal digits.
These leave ample opportunity for a fortunate reader with a talent for proof.

\newpage 

\subsection{Arena}

{\em Quantum field theorists} predict
probabilities for outcomes of interactions of particles, as at the large hadron collider, by
making expansions in coupling constants that are hopefully small enough
to permit accurate comparison with experimental data, as is the case for the standard model of particle
physics, at high energy. 

They do this by evaluating integrals defined by Feynman diagrams, whose 
external edges specify the process under study, while the internal edges represent
the possibility of so-called virtual propagation, in which the energy $E$ and momentum
$p$ are no longer related by Einstein's mass-shell condition $E^2=(mc^2)^2+(pc)^2$,
for a particle of mass $m$, with $c$ the speed of light. In a spacetime with even dimension 
$D$, like our own with $D=4$, the propagator, between spacetime events at adjacent
vertices of the diagram, involves a {\em Bessel function}, when $m>0$.
Integrations must be performed over all spacetime separations of the vertices. 

Thus, whenever the diagram has at least two internal massive edges, it yields an integral
of a product of Bessel functions.  That may be disguised,
by Fourier transformation, or by use of parametric integration, yet it is always the case.
I have therefore devoted considerable effort to the study of integrals of Bessel functions, notably
in work with David Bailey, Jon Borwein and Larry Glasser~\cite{BBBG}.

{\em Algebraic geometers} may take interest in a Feynman integral when shown its equivalent
representation as a projective integral over Schwinger parameters, with an integrand
whose denominator is a 
polynomial in the parameters associated to the edges, with coefficients determined by
the external data and the internal masses. From this it may be possible to determine,
by what Spencer Bloch likes to call {\em pure thought}, the type of number that
the Feynman diagram will yield and in particular whether the result will
be a simple polylogarithm or something more refined, such as the elliptic
dilogarithms in~\cite{ABW,BKV2} or the elliptic trilogarithm found by 
Bloch, Kerr and Vanhove in~\cite{BKV} for a sunrise diagram, with off-shell
external data, in two spacetime dimensions. This has 5 Bessel functions, at 3 loops, giving
\begin{equation}
{\cal S}_{5,3}(w):=2^3\int_0^\infty I_0(wt)K_0^4(t)\,t\,dt.
\label{S53w}\end{equation}
In 2007, I had conjectured~\cite{Bnome} that
\begin{equation}
S_{5,3}:={\cal S}_{5,3}(1)=
\frac{\pi^3}{2}\left(1-\frac{1}{\sqrt{5}}\right)
\left(\sum_{n=-\infty}^\infty e^{-n^2\pi\sqrt{15}}\right)^4
\label{S53nome}\end{equation}
in the on-shell case, with $w=1$. This is now proven~\cite{BKV,S}.

{\em Number theorists} may take interest in a Feynman integral when shown its evaluation, 
as in~(\ref{S53nome}).
If the coefficients of the polynomial of Schwinger parameters are rational numbers
the integral is, by definition, a {\em period}~\cite{KZ}. This arises when there
is a single large energy scale in the process, allowing effective
neglect of internal masses. Then the period  may be a multiple zeta value (MZV) as in my
work with Dirk Kreimer~\cite{BK}.
MZVs hold lively interest for mathematicians~\cite{GKZ,IKZ}
and I have devoted considerable effort to their study, notably
in work with Johannes Bl\"{u}mlein and Jos Vermaseren~\cite{BBV}.
Yet Francis Brown and Oliver Schnetz have shown that
even single-scale massless diagrams outgrow the kindergarten of polylogarithms,
such as MZVs,
when the loop-number (which mathematicians call the first Betti number) of the diagram
reaches 7, for massless diagrams with two external particles~\cite{BSK3}. 

When the scale is set by a single mass, results may be given by L-series
of modular forms, as in~(\ref{S53nome}), where the 3-loop evaluation
\begin{equation}
\frac{S_{5,3}}{\zeta(2)}=\frac{48}{5}L_5(1)
\label{S53L1}\end{equation}
comes from combining work in~\cite{BKV} with work by Mathew Rogers, James Wan and 
John Zucker~\cite{RWZ} on an L-series, $L_5(s)$,
defined by a modular form~\cite{PTV} of weight 3 and level 15,
with a functional equation derived from Kloosterman moments at $n=5$.
This is the subject of Sections~5.1 and~7.1.

My purpose is to address the {\em simplest} cases in which
interests of quantum field theorists, algebraic geometers and number theorists
intersect and the Feynman integrals are not mere polylogarithms.

Accordingly, let $D=2$, since in two spacetime dimensions I need 
not instruct the reader in Freeman Dyson's miracle of renormalization~\cite{Dyson}, 
required to render finite the Feynman integrals of the real world, with $D=4$.

Let all particles be massive and spin-less, as is the Higgs boson. 
Crucially, let them have the same mass, $m$, and work in a system of units where
$m=c=1$ and Planck's constant is $h=2\pi$. 
Then an internal edge has a propagator $K_0(t)$,
were $t$ is the separation, in  proper time, of the vertices that the edge connects.
The Bessel function $K_0(t)$ falls off exponentially fast at large $t$,
so there will be no infra-red problems. At $t=0$ it has a logarithmic
singularity, benign enough to avoid ultra-violet problems.

Finally, and most drastically, let each diagram have precisely two vertices.
Then, as illustrated in Section~7, there are two cases: 
vacuum diagrams, with no external data, or so-called sunrise diagrams,
where energy $E$ and momentum $p$ enter at one vertex and leave
at the other. For the latter, work on the external mass-shell, with $w^2:=E^2-p^2=1$.
Then the external data will contribute $I_0(t)$ to the integrand.
The Bessel functions $I_0(t)$ and $K_0(t)$ are born of the same second order differential
equation, but $I_0(t)$ is quite contrary to its sibling, being well behaved at $t=0$, where it is regular,
and poorly behaved at large $t$, where it increases exponentially. 
 
Thus we shall be studying single integrals of products of Bessel functions.
This tiny part of the vast arena of quantum field theory holds remarkable surprises,
illuminated by L-series associated with Kloosterman sums.

\subsection{{\em Pr\'{e}cis} of Section~2, on Kloosterman sums}%1.2

Here, I define {\em Kloosterman sums} in finite fields, ${\bf F}_q$, with $q=p^k$, for
prime $p\geq2$ and integer exponent $k\geq1$. These sums bring us as close
as possible to emulating Bessel functions, while
rigourously eschewing all branches of analysis. 
From them I shall obtain
what Phillipe Robba (1941--1988) called {\em symmetric powers of the $p$-adic Bessel equation}~\cite{R}.
These provide integers, $c_n(q)$, for integer powers $n\geq1$.  I shall refer to
them as {\em Bessel moments} in finite fields. They may seem remote
from the hard analysis required for Feynman diagrams with masses, which involves integration 
over products of Bessel functions that are themselves defined as integrals
or infinite sums. Yet $c_n(q)$ brings us as close as possible to capturing the features of an integral 
of the product of $n$ Bessel functions while allowing ourselves only finite summations
that yield integers.
They key fact is that $c_n(p^k)$ is {\em predictable} for $k>d(p,n)$,
where $d(p,n)$ is a vital integer, associated with a functional equation for
Robba's $p$-adic problem. This has deep consequences 
for the evaluation of the Feynman integrals.

\subsection{{\em Pr\'{e}cis}  of Section~3, with Algorithm~1}%1.3

This is devoted to determining data for a generating function
that gives important information about all moments $c_n(p^k)$, for each prime $p$.
I have tried to keep this section simple and self-contained,
avoiding profound mathematical concepts that may be unfamiliar to physicists, like myself.
Yet it is still a delicate task, since one must carefully distinguish the
cases of odd and even $n$, with an important sign, $s(p,n)$, prominent for
odd $n$, where the residue of $p$ modulo 8 is also important. Moreover,
the prime $p=2$ is a very special case that was only recently understood,
thanks to fine work by Zhiwei Yun~\cite{YV}, whose crucial contribution I have
condensed to a single line, in~(\ref{abrec2}), where the even prime, 2,
talks deeply with its neighbour, 3.

A result for the sign, from Lei Fu and Daqing Wan~\cite{FW3}, has been
condensed to a simple recursion in~(\ref{srec}), from which I derive Algorithm~1,
enabling efficient  computation of a generating function for $d(p,n)$.
Fu and Wan also corrected a faulty conjecture by Robba on 
the case $p\equiv3$ mod 4 in~(\ref{bgen}).

\subsection{{\em Pr\'{e}cis}  of Section~4, with a theorem and corollaries}%1.4

A reader more concerned with results than with derivations may wish to skim
the details of the previous section and concentrate instead on this section, where 
Theorem~1 encapsulates all the hard-won findings on $d(p,n)$  and leads to Corollary~1, 
with a remark on an uncanny parallel to my work with Kreimer~\cite{BK}, which connects
enumerations of irreducible MZVs to enumeration of modular forms. Corollary~2
is included for those who prefer floors to generating functions, with a remark
that favours the latter.

\subsection{{\em Pr\'{e}cis}  of Section~5, with Conjectures~1, 2 and 3}%1.5

Here, I propose Conjecture~1, which constrains
Bessel moments, with odd $n$, even more tightly than the functional equation.
It is proven at $n\leq5$ and agrees well with Ronald Evans' inspired discoveries~\cite{CE,E}
at $n=7$. I set store on it to illuminate uncharted territory at $n\geq9$, where
it has survived tenacious testing. Moreover, I refine it by Conjectures~2 and~3,
which distinguish the cases $n\equiv1$ mod 4 and $n\equiv3$ mod 4, where 
they are even more predictive. They too survive deep testing.

\subsection{{\em Pr\'{e}cis}  of Section~6, with Algorithms~2 and~3}%1.6

Here, I hone the computational challenges to their
ineluctable essentials. This was a necessary step, since the
innermost declarations of Algorithm~2 were executed 
about $10^{13}$ times, in tests of the conjectures. 
Algorithm~3 ensures that one uses Algorithm~2 as sparingly
as possible.

\subsection{{\em Pr\'{e}cis}  of Section~7, with Conjectures~4, 5, 6 and 7}%1.7

Here, at last, we get to study the Feynman integrals, finding wonderfully
simple results for them as critical values of L-series defined by infinite products
over data at the primes, from Kloosterman moments. 
Some of these results were reported in~\cite{BLinz,BS}.
Since then there has been dramatic progress, thanks to
work with Anton Mellit~\cite{BM}. We have been able to achieve 
results at $n=7$, where the functional equation was found to
be significantly different. By using {\em determinants}
of Feynman integrals, including those from vacuum diagrams,
we reached non-critical values that are
the subject of conjectures by Fernando Rodriguez Villegas,
on logarithmic {\em Mahler measures}~\cite{V}. Moreover we reach a place
that was inaccessible to him: the value of a L-series with modular weight
6 at $s=7$. This we claim to evaluate in Conjecture~6 of Section~7.8,
using a $2\times2$ determinant that includes 7-loop Feynman integrals
with 8 Bessel functions.

The route to Conjecture~6 was via the striking Conjecture~5 
of Section~7.7, which is dedicated to the memory of
Richard Crandall, in regret that he did not live to see it.
Further determinants yield powers of $\pi$ and, from diligent study of these,
I advance Conjectures~4 and~6, with examples in~(\ref{det15}) and~(\ref{det16}),
by way of {\em amuse bouche}. These have been checked numerically by evaluating
hundreds of Feynman integrals, with up to 30 loops, at 500-digit precision.

\subsection{{\em Pr\'{e}cis}  of Section~8, on harder problems}%1.8

{\em What proverb more common, what proverb more true, than that 
after pride comes a fall?} 

Mindful of Kingsley's~\cite{K} question, I report attempts with $n>8$ Bessel functions.
At $n=9$, Conjecture~2 places hitherto unexplored
constraints on the data at the primes. In a sense that I make precise, 
$c_9(p)$ determines $c_9(p^2)$ {\em more often than not}, for $p\leq631$.
Yet that has not produced, so far, a result for Feynman integrals with 9 Bessel functions.
As Francis Brown is fond of reminding me, all good things come to an end.
I respond with a question from his almost namesake, Browning~\cite{Br}.

{\em Ah, but a man's reach should exceed his grasp, or what's a heaven for?}

\newpage

\section{Functional equation for Kloosterman moments}%2

Kloosterman sums, $K(a)$, with $a\in{\bf F}_q$ and $q=p^k$, are defined by
\begin{equation}
K(a):=\sum_{x\in{\bf F}_q^*}\exp\left(\frac{2\pi i}{p}\trace\left(x+\frac{a}{x}\right)\right)
\label{K}\end{equation}
with a trace of Frobenius, whose details I postpone to Section~6, were they are truly needed.
The key information that we need for Feynman integrals
is given by the Bessel moments $c_n(q)$, defined by
\begin{equation}
c_n(q):=-\frac{1+S_n(q)}{q^2},\quad
S_n(q):=\sum_{a\in{\bf F}_q^*}\sum_{j=0}^n\left[g(a)\right]^j\left[h(a)\right]^{n-j}
\label{c}\end{equation}
with $K(a)=-g(a)-h(a)$ and $g(a)h(a)=q$. These moments are integers
that bring us as close as possible to emulating Feynman integrals, while performing
only finite summations. Hence I have devoted considerable effort to refining
algorithms for their efficient computation.
  
Notwithstanding the appearance of an infinite sum in the exponential of
\begin{equation}
Z_n(p,T):=\exp\left(-\sum_{k>0}\frac{c_n(p^k)}{k}T^k\right)
\label{Z}\end{equation}
the result is a polynomial in $T$, whose degree, $r(p,n)$, I shall call
the {\em raw} degree. In essence,~(\ref{Z}) tells us that $c_n(p^k)$ is
predictable for $k>r(p,n)$.

For $n\leq4$ the situation is very simple,
with $Z_1(p,T)=Z_2(p,T)=1$,
\begin{equation}
Z_3(p,T)=1-\left(\frac{p}{3}\right)T,\quad
Z_4(p,T)=\left\{\begin{array}{ll}
1&\mbox{if }p=2\\
1-T&\mbox{if }p>2\end{array}\right.
\label{Z34}\end{equation}
and a Legendre symbol $\left(\frac{p}{3}\right)=0\mbox{ or }\pm1$, for 
$p\equiv0\mbox{ or }\pm1$ mod 3.

It is often possible to improve predictability by extracting factors from~(\ref{Z}).
Let $y=p^{k-1}T$, for odd $n=2k+1$, and $y=p^{k-2}T$, for even $n=2k$.
Then there are exponents $a(p,n)$ and $b(p,n)$ such that
\begin{equation}Z_n(p,T)=(1-y)^{a(p,n)}(1+y)^{b(p,n)}M_n(p,T)
\label{M}\end{equation}
where $M_n(p,T)$ has even degree $2d(p,n)=r(p,n)-a(p,n)-b(p,n)$,
which I shall call the {\em net} degree. Most importantly, there is a functional equation
\begin{equation}
M_n(p,T)=\left(p^{n-3}T^2\right)^{d(p,n)}M_n\left(p,\frac{1}{p^{n-3}T}\right)
\label{Msym}\end{equation}
from which it follows that $Z_n(p,T)$ may be computed from moments of
Kloosterman
sums in finite fields ${\bf F}_q$ with characteristic $p$ and $q\leq p^{d(p,n)}$,
provided that one knows the raw degree of~(\ref{Z}) and the
exponents in~(\ref{M}). It is proven that $d(p,n)\leq1$ for $n\leq8$. For
$5\leq n\leq8$, we may therefore determine $c_n(p^k)$ for all $k>1$ knowing only $c_n(p)$.
This circumstance is associated with the existence of automorphic
forms of modular weight $n-2$ for $5\leq n\leq8$, where the L-series associated 
with $c_n(p)$ provide evaluations of Feynman integrals in quantum 
field theory.

\section{Case by case study of moments}%3

We must take care to distinguish cases by the parities of $n$ and $p$, as follows.

\subsection{Moments with odd $n$ and $p>2$}%3.1

Let $p>2$ be an odd prime and $n$ be an odd integer. Then the raw degree is given
recursively  by
\begin{equation}
r(p,n)=r(p,n-2)+\left\{\begin{array}{rl}
0&\mbox{if } p|n\\1&\mbox{otherwise}\end{array}\right.
\label{rrec}\end{equation}
with $r(p,1)=0$.  Using {\em floor} delimiters in
$\left\lfloor x\right\rfloor$ to denote the largest integer not exceeding $x$,
we obtain the solution~\cite{R}
\begin{equation}
r(p,n)=\frac{n-1}{2}-\left\lfloor\frac{n+p}{2p}\right\rfloor.
\label{rodd}\end{equation}
The {\em sign}~\cite{FW3} that we need is given by $s(p,1)=1$ and the recursion
\begin{equation}
\frac{s(p,n)}{s(p,n-2)}=t(p,n):=\left\{\begin{array}{rl}
\left(\frac{-2}{p}\right)&\mbox{if } p|n\\\left(\frac{p}{n}\right)&\mbox{otherwise}\end{array}\right.
\label{srec}\end{equation}
with the Legendre symbol
\begin{equation}
\left(\frac{-2}{p}\right)=(-1)^{(p-1)(p-3)/8}=\left\{\begin{array}{rl}
+1&\mbox{if }p\equiv1\mbox{ or }3\mbox{ mod }8\\
-1&\mbox{if }p\equiv5\mbox{ or }7\mbox{ mod }8\end{array}\right.
\label{leg}\end{equation}
occurring when $p$ divides $n$. Otherwise we encounter the Jacobi symbol
\begin{equation}
\left(\frac{p}{n}\right)=\left(\frac{p}{-n}\right)=\left(\frac{n}{p}\right)\left(\frac{-1}{p}\right)^{(n-1)/2}
=\left(\frac{p}{n+2p}\right)\left(\frac{-1}{p}\right)=\left(\frac{p}{n+4p}\right)
\label{jac}.\end{equation}
In both cases, $t(p,n)=t(p,-n)=t(p,n+4p)$ and we easily evaluate
\begin{equation}
\frac{s(p,n)}{s(p,-2-n)}=\prod_{k=0}^{n}t(p,n-2k)t(p,2k-n)=1.
\label{smin}\end{equation}
More care must be taken to determine the sign of
\begin{equation}
\frac{s(p,n+2p)}{s(p,-2-n)}=t(p,p)\prod_{k=0}^{(n+p)/2}t(p,p+2k)t(p,p-2k)=
\left(\frac{-2}{p}\right)\left(\frac{-1}{p}\right)^{u(p,n)}
\label{scare}\end{equation}
with an exponent
\begin{equation}
u(p,n):=\frac{n+p}{2}-\left\lfloor\frac{n+p}{2p}\right\rfloor=\frac{p+1}{2}+r(p,n).
\label{sexp}\end{equation}
Noting that $\left(\frac{-1}{p}\right)^{(p+1)/2}=(-1)^{(p^2-1)/4}=1$, we obtain, from~(\ref{smin},\ref{scare}), 
\begin{equation}
s(p,2p-2-n)=s(p,n+2p)=\left(\frac{-2}{p}\right)\left(\frac{-1}{p}\right)^{r(p,n)}s(p,n).
\label{shalf}\end{equation}
Moreover, $s(p,n)=s(p,n+4p)$, since $r(p,n+2p)-r(p,n)=p-1$ is even. 
It follows from~(\ref{shalf}) that the signs $s(p,n)$ with $1<n<p$ suffice
to determine all other signs.

The net degree $2d(p,n)=r(p,n)-a(p,n)-b(p,n)$ 
is obtained by removing a factor
$(1-y)^{a(p,n)}(1+y)^{b(p,n)}$, with $y=p^{(n-3)/2}T$ and exponents  
\begin{equation}
a(p,n)=\frac{1-s(p,n)(-1)^{r(p,n)}}{2},\quad b(p,n)=\frac{1-s(p,n)}{2}
\label{abodd}\end{equation}
that are either 0 or 1. It follows that 
\begin{equation}
\ep(p,n):=\left\lfloor\frac{r(p,n)}{2}\right\rfloor-d(p,n)
=\frac{1-s(p,n)}{2}\,\frac{1+(-1)^{r(p,n)}}{2}
\label{eps}\end{equation}
vanishes when $r(p,n)$ is odd or $s(p,n)=1$. 
Sufficient information is encoded by the economical polynomial
\begin{equation}
g_p(x):=\sum_{p>2k+1>0}\ep(p,2k+1)x^{2k+1}
\label{gp}\end{equation}
for which $\left\lfloor\frac{p-3}{4}\right\rfloor$
Legendre symbols suffice, in the following algorithm.

{\bf Algorithm 1.} [Compute $g_p(x)$ for prime $p$.]
Set $s=n=1$ and $g=0$. While $n+4<p$, add 4 to $n$,
change the sign of $s$ if $n(2-n)$ is not a square modulo $p$,
add $x^n$ to $g$ if $s$ is negative. Return $g_p(x)=g$.

For convenience, here are the results for odd primes $p\leq31$:
\begin{equation}
g_3=g_5=0,\quad
g_7=g_{13}=x^5,\quad
g_{11}=g_{29}=x^5+x^9,\quad
g_{17}=x^9,
\label{gs}\end{equation}
\begin{equation}
g_{19}=x^9+x^{13}+x^{17},\quad
g_{23}=x^{17},\quad
g_{31}=x^9+x^{17}+x^{21}+x^{25}.
\label{gl}\end{equation}

To determine $\ep(p,n)$ for odd $n\ge p$, we use~(\ref{shalf}), which gives
\begin{equation}
\ep(p,2p-2-n)=\ep(p,n+2p)=\left\{\begin{array}{rl}
\ep(p,n)&\mbox{if }\left(\frac{-2}{p}\right)=+1\\
1-\ep(p,n)&\mbox{if }\left(\frac{-2}{p}\right)=-1.\end{array}\right.
\label{ehalf}\end{equation}
Defining the polynomial
\begin{equation}
h_p(x):=
\left(1+\left(\frac{-2}{p}\right)x^{2p}\right)\left(g_p(x)+\left(\frac{-2}{p}\right)x^{2p-2}g_p(1/x)\right)
\label{hh}\end{equation}
we obtain, from~(\ref{gp},\ref{ehalf}) and the periodicity of $\ep(p,n)=\ep(p,n+4p)$,
\begin{equation}
(1-x^{4p})\sum_{k\geq0}\ep(p,2k+1)x^{2k+1}=
\left\{\begin{array}{rl}h_p(x)&\mbox{if }\left(\frac{-2}{p}\right)=+1\\
e_p(x)+h_p(x)&\mbox{if }\left(\frac{-2}{p}\right)=-1\end{array}\right.
\label{egen}\end{equation}
with an extra polynomial 
\begin{equation}
e_p(x):=\sum_{p\leq4k+3<3p}x^{4k+3}=\left\{\begin{array}{rl}
(x^{p+2}-x^{3p})/(1-x^4)&\mbox{if }p\equiv5\mbox{ mod }8\\
(x^p-x^{3p+2})/(1-x^4)&\mbox{if }p\equiv7\mbox{ mod }8\end{array}\right.
\label{ee}\end{equation}
to take account of the cases with $\left(\frac{-2}{p}\right)=-1$ in~(\ref{ehalf}).

To determine a generating function $\sum_{k\geq0}d(p,2k+1)x^{2k+1}$
from~(\ref{eps},\ref{egen}), it suffices to evaluate the polynomial
\begin{equation}
f_p(x):=(1-x^2)(1-x^{4p})\sum_{k\geq0}\left\lfloor\frac{r(p,2k+1)}{2}\right\rfloor x^{2k+1}
\label{fdef}\end{equation}
which depends on the sign $\left(\frac{-1}{p}\right)=(-1)^{(p-1)/2}$, as follows:
\begin{equation}
f_p(x)=x^5\left(\frac{1-x^{4p}}{1-x^4}\right)-
\left\{\begin{array}{cl}
x^p&\mbox{if }p\equiv1\mbox{ mod }4\\
x^{p+2}&\mbox{if }p\equiv3\mbox{ mod }4.\end{array}\right.
\label{ff}\end{equation}
Before assembling these results for odd $n$, we consider the even moments.

\subsection{Moments for even $n$ and $p>2$}%3.2

Let $p>2$ be an odd prime and $n=2k$ be an even positive integer. 
Then the raw degree is given recursively  by~(\ref{rrec})
with $r(p,2)=0$. The solution is $r(p,2k)=k-1-\lfloor\frac{k}{p}\rfloor$.
The net degree $2d(p,n)=r(p,n)-a(p,n)-b(p,n)$ 
is obtained by removing a factor
$(1-y)^{a(p,n)}(1+y)^{b(p,n)}$, with $y=p^{(n-4)/2}T$ and exponents
determined by~\cite{FW1,FW2,FW3}
\begin{eqnarray}
\sum_{k>0}\left(a(p,2k)+b(p,2k)\right)x^{2k}&=&
\frac{x^4}{1-x^4}+\frac{x^{2p}}{(1-x^2)(1-x^{2p})}\label{abgen}\\
(1-x^2)(1-x^{4p})\sum_{k>0}b(p,2k)x^{2k}&=&
\left\{\begin{array}{rl}
0&\mbox{if }p\equiv1\mbox{ mod 4}\\
x^{2p}&\mbox{if }p\equiv3\mbox{ mod 4}\end{array}\right.\label{bgen}
\end{eqnarray}
with a distinction of cases in~(\ref{bgen}) that does not affect
the outcome for~\cite{R}
\begin{equation}
d(p,2k)=\left\lfloor\frac{k-1}{2}\right\rfloor-\left\lfloor\frac{k}{p}\right\rfloor.
\label{deven}\end{equation} 
 
\subsection{Moments for odd $n$ at $p=2$}%3.3
 
Here one uses the recursions~(\ref{rrec},\ref{srec}),
where there is now no need to consider the case $p|n$.
Hence we obtain a raw degree $r(2,2k+1)=k$
and an easily evaluated Jacobi symbol
\begin{equation}
\frac{s(2,n)}{s(2,n-2)}=\left(\frac{2}{n}\right)=(-1)^{(n^2-1)/8}=
\left\{\begin{array}{rl}
+1&\mbox{if }n\equiv1\mbox{ or }7\mbox{ mod }8\\
-1&\mbox{if }n\equiv3\mbox{ or }5\mbox{ mod }8\end{array}\right.
\label{srec2}\end{equation}
in the recursion for the sign. Then, by induction, we prove that
\begin{equation}
s(2,n)=(-1)^{(n+3)(n^2-1)/48}=\left\{\begin{array}{rl}
+1&\mbox{if }n\equiv1\mbox{, 5 or 7 mod 8}\\
-1&\mbox{if }n\equiv3\mbox{ mod }8\end{array}\right.
\label{ssol2}\end{equation}
with $p=2$ singling out the residue 3 modulo 8 as a special case. 
From~(\ref{abodd}), we obtain
\begin{equation}
\sum_{k\geq0}a(2,2k+1)x^{2k+1}=\frac{x^7}{1-x^8},\quad
\sum_{k\geq0}b(2,2k+1)x^{2k+1}=\frac{x^3}{1-x^8}.
\label{abodd2}\end{equation}

\subsection{Moments for even $n$ at $p=2$}%3.4

The subtlest question has been left to last. At $p=2$, what is the pattern of exponents in~(\ref{M})
for even moments?
The raw degree is $r(2,2k)=\left\lfloor\frac{k-1}{2}\right\rfloor$,
but to obtain the net degree $2d(2,2k)=r(2,2k)-a(2,2k)-b(2,2k)$
we need a pair of recursions recently proven by Yun~\cite{YV}:
\begin{equation}
a(2,2k+24)-a(2,2k)=1=b(2,2k+24)-b(2,2k)
\label{abrec2}\end{equation}
with 24 signally a dialogue between the neighbours $p=2$ and $p=3$.

To determine $a(2,2k)$ and $b(2,2k)$ for $2k\leq24$, it suffices
to evaluate $Z_{2k}(2,T)$ from moments  of Kloosterman sums in ${\bf F}_{2^N}$ 
with $N\leq5$, which is easily accomplished by Algorithm 2. 
We obtain $Z_2(2,T)=Z_4(2,T)=1$, $Z_6(2,T)=1+y$ and $Z_8(2,T)=1-y$
merely from working in ${\bf F}_2$. Work in ${\bf F}_4$ 
quickly determines $Z_{10}(2,T)=1+6T+2^7T^2$ and $Z_{12}(2,T)=1-y^2$,
with $y=2^{k-2}T$ for $Z_{2k}(2,T)$. Thereafter
\begin{eqnarray}
Z_{14}(2,T)&=&(1+y)(1-54T+2^{11}T^2)\label{Z14}\\
Z_{16}(2,T)&=&(1-y)(1+132T+2^{13}T^2)\label{16}\\
Z_{18}(2,T)&=&(1-y^2)(1-114T+2^{15}T^2)\label{18}\\
Z_{20}(2,T)&=&(1-y^2)(1+72T+2^{17}T^2)\label{20}
\end{eqnarray}
reveal that $d(2,2k)\leq1$ for $2k\leq20$. Moments in ${\bf F}_{32}$ complete the picture:
\begin{eqnarray}
Z_{22}(2,T)&=&(1+y)(1-270(T+2^{19}T^3)-230720T^2+2^{38}T^4)\label{Z22}\\
Z_{24}(2,T)&=&(1-y)(1-y^2)(1+12T+2^{21}T^2).\label{Z24}
\end{eqnarray}
The values $a(2,2k)$ and $b(2,2k)$, with $k=1\mbox{ to }12$, thus form the sequences
\begin{equation}\begin{array}{llllllllllll}
0,&0,&0,&1,&0,&1,&0,&1,&1,&1,&0,&2\\
0,&0,&1,&0,&0,&1,&1,&0,&1,&1,&1,&1\end{array}
\label{abseq}\end{equation}
respectively. Then the recursions in~(\ref{abrec2}) prove that
\begin{eqnarray}
\sum_{k>0}a(2,2k)x^{2k}&=&\frac{x^8}{1-x^8}\left(1+\frac{x^4}{1-x^6}\right)\label{aeven2}\\
\sum_{k>0}b(2,2k)x^{2k}&=&\frac{x^6}{(1-x^6)(1-x^8)}.\label{beven2}
\end{eqnarray}
Hence we obtain, from~(\ref{abodd2},\ref{aeven2},\ref{beven2}), the generating function
\begin{equation}
\sum_{n>0}d(2,n)x^n=\frac{x^5}{(1-x^2)(1-x^4)}+\frac{x^{10}}{(1-x^4)(1-x^6)}.
\label{dboth2}\end{equation}

\newpage

\section{Generating functions for all cases}%4

I now assemble the results for $d(p,n)$. For prime $p>2$, let
\begin{eqnarray}
d_p(x)&:=&\left\{\begin{array}{rl}
x^p(1+x^{2p})&\mbox{if }p\equiv1\mbox{ mod }8\\
x^{p+2}(1+x^{2p})&\mbox{if }p\equiv3\mbox{ mod }8\\
x^p(1+x^2)&\mbox{if }p\equiv5\mbox{ or }7\mbox{ mod }8\end{array}\right.\label{dd}\\
C_p(x)&:=&\frac{x^5}{(1-x)(1-x^4)}-\frac{x^{2p}}{(1-x^2)(1-x^{2p})}\label{CC}\\
D_p(x)&:=&C_p(x)
-\frac{d_p(x)}{(1-x^4)(1-x^{4p})}
-\frac{g_p(x)+\left(\frac{-2}{p}\right)x^{2p-2}g_p(1/x)}{1-\left(\frac{-2}{p}\right)x^{2p}}\label{DD}
\end{eqnarray}
with $g_p(x)$ determined by $\left\lfloor\frac{p-3}{4}\right\rfloor$ Legendre symbols, via Algorithm 1.
 
{\bf Theorem 1.} The net degree $2d(p,n)$ is generated by
\begin{equation}
\sum_{n>0}d(p,n)x^n=
\left\{\begin{array}{rl}C_3(x)&\mbox{if }p=2\\D_p(x)&\mbox{if }p>2.\end{array}\right.
\label{th1}\end{equation}

{\bf Proof.} For odd $n$,~(\ref{ff}) 
gives $d_p(x)=x^p(1+x^{2p})$ for $p\equiv1\mbox{ mod }8$ and 
$d_p(x)=x^{p+2}(1+x^{2p})$ for $p\equiv3\mbox{ mod }8$. For
$p\equiv5\mbox{ mod }8$, the extra term in~(\ref{ee}) gives
$d_p(x)=x^p(1+x^{2p})+x^{p+2}-x^{3p}=x^p(1+x^2)$.
For $p\equiv7\mbox{ mod }8$, it gives
$d_p(x)=x^{p+2}(1+x^{2p})+x^{p}-x^{3p+2}=x^p(1+x^2)$.
Since $d_p(x)$ and $g_p(x)$ are odd functions,
only $C_p(x)$ contributes for even $n$ and $p>2$, 
giving~(\ref{deven}). Setting $p=3$ in~(\ref{CC}), we obtain 
the result for $p=2$ in~(\ref{dboth2}).~$\square$

{\bf Corollary 1.}  Let $M(w)$ be the dimension of the space of cusp forms of weight $w$ for the
fundamental modular group. Then
\begin{equation}
d(2,2k)=d(3,2k)=M(2k-2)=\left\lfloor\frac{k-1}{2}\right\rfloor-\left\lfloor\frac{k}{3}\right\rfloor.
\label{cusp}\end{equation}

{\bf Proof.} The cusp forms are by generated by products of the unique cusp form with
$w=12$ and powers of Eisenstein series with weight 4 or 6. Hence
\begin{equation}
\sum_{w>0}M(w)x^w=\frac{x^{12}}{(1-x^4)(1-x^6)}.
\label{Mw}\end{equation}
The even term in~(\ref{dboth2}) proves that $M(2k-2)=d(2,2k)$.
Theorem~1 gives $d(2,2k)=d(3,2k)$. Then~(\ref{deven}), at $p=3$, completes 
the proof of~(\ref{cusp}).~$\square$

{\bf Remark 1.} The Broadhurst-Kreimer conjectures~\cite{BCan,BK}, for the embedding  
of the two-letter alphabet of MZVs in the three-letter alphabet~\cite{BBV}
of nested alternating sums, also use the generating function~(\ref{Mw}) for cusp forms.
I was reluctant to regard this as anything more than coincidence,
since it was hard for me to see how MZVs, at genus zero, talk to Eisenstein series,
at genus 1. Yet Don Zagier~\cite{GKZ,IKZ} and Francis Brown~\cite{Brown} 
seem to believe this to be no accident.
In the present case, the coincidences of Corollary~1 are proven.

{\bf Corollary 2.}  $d(3,2k+1)=\left\lfloor\frac{k}{3}\right\rfloor$ and 
$d(5,2k+1)=\left\lfloor\frac{k-2}{2}\right\rfloor-\left\lfloor\frac{k-2}{10}\right\rfloor$.

{\bf Proof.}  Noting that $g_3(x)=g_5(x)=0$, we obtain
\begin{eqnarray}
\sum_{k\geq0}d(3,2k+1)x^{2k+1}&=&\frac{x^7}{(1-x^2)(1-x^6)}\label{floor3}\\
\sum_{k\geq0}d(5,2k+1)x^{2k+1}&=&\frac{x^9}{(1-x^2)(1-x^4)}-\frac{x^{25}}{(1-x^2)(1-x^{20})}\label{floor5}
\end{eqnarray}
from~(\ref{DD}). Conversion to the stated floors follows immediately.~$\square$

{\bf Remark 2.} For $p>5$, floors may be obtained by collecting terms, as here:
\begin{equation}
\sum_{k\ge0}d(7,2k+1)x^{2k+1}=
\frac{x^7 + x^{13} + x^{17} + x^{19} - x^{21} + x^{23} + x^{25} + x^{29}}{(1-x^2)(1-x^{28})}
\label{n7}\end{equation}
\begin{equation}
\sum_{k\ge0}d(17,2k+1)x^{2k+1}=
\frac{x^5 + x^{11} + x^{13} + x^{19} + x^{25} + x^{27} + x^{31} + x^{35}}{(1-x^2)(1-x^{34})}
\label{n17}\end{equation}
with each monomial in the numerator yielding a floor. In these examples, we obtain 8 terms,
while there is only one in $g_7(x)=x^5$, or $g_{17}(x)=x^9$. In general,
the number of monomials in the numerator is given by
\begin{equation}
n_p:=\left\{\begin{array}{cl}\frac{p-1}{2}&\mbox{if }p\equiv1\mbox{ or 3 mod 8}\\
p-1&\mbox{if }p\equiv5\mbox{ mod 8}\\
p+1&\mbox{if }p\equiv7\mbox{ mod 8}\end{array}\right.
\label{np}\end{equation}
while the number of monomials in $g_p(x)$ is asymptotic to $\frac{1}{8}p$,
providing a saving in data  by factors of 4 or 8, at large $p$.

\section{Cores from odd moments}%5

The {\em core} of a non-zero integer $N$ is the unique square-free integer
$d|N$ such that $N/d$ is the square of an integer. For odd $n$, let
\begin{eqnarray}
c^\pm_n(p)&:=&p^{m(n)}Z_n\left(p,\frac{\pm1}{p^{(n-3)/2}}\right)\label{cpm}\\
m(n)&:=&\left\{\begin{array}{cl}
(n-1)^2/16&\mbox{if }n\equiv1\mbox{ mod 4}\\
(n-3)(n+1)/16&\mbox{if }n\equiv3\mbox{ mod 4}.\end{array}\right.\label{mpm}
\end{eqnarray}

{\bf Conjecture 1.} For odd $n\ge1$ and prime $p\ge2$, the constants $c^\pm_n(p)$ are 
non-negative integers. Moreover, if $c^\pm_n(p)$ is non-zero, then its core has no 
prime divisor greater than $n$.

{\bf Remark.} Recalling that $y:=p^{(n-3)/2}T$, for odd $n$, we see from~(\ref{M}) that
$c_n^+(p)=0$ if $a(p,n)>0$ and $c_n^-(p)=0$ if $b(p,n)>0$. Moreover  
$c_1^\pm(p)=1$ and $c_3^\pm(p)=1\mp\left(\frac{p}{3}\right)$ satisfy
Conjecture~1. The conjecture has highly non-trivial content for odd $n\ge5$.

\subsection{A modular form at $n=5$}%5.1

At $n=5$, with $y=pT$, we have $Z_5(3,T)=1+y$, $Z_5(5,T)=1-y$
and $Z_5(p,T)=1-y^2$ for $\left(\frac{p}{15}\right)=-1$. Since
$c_5^\pm(p):=p Z_5(p,\pm1/p)$ is evaluated at $y=\pm1$,
Conjecture~1 holds in all of these cases. 

When $\left(\frac{p}{15}\right)=1$, we have $Z_5(p,T)=1-c_5(p)T+y^2$
and hence obtain $c_5^\pm(p)=2p\mp c_5(p)$. It follows that 
$4p=c_5^+(p)+c_5^-(p)$ and $-2c_5(p)=c_5^+(p)-c_5^-(p)$.
There are {\em unique} positive integers, $u$ and $v$, such that
\begin{equation}
4p=\left\{\begin{array}{cl}
15u^2+v^2&\mbox{if }\left(\frac{p}{3}\right)=\left(\frac{p}{5}\right)=+1\\
3u^2+5v^2&\mbox{if }\left(\frac{p}{3}\right)=\left(\frac{p}{5}\right)=-1\end{array}\right.
\label{Z5}\end{equation}
and it follows from work by Peters, Top and van der Vlugt~\cite{PTV}  
that the pair $(u,v)$ determines
\begin{equation}
-2c_5(p)=\left\{\begin{array}{cl}
15u^2-v^2&\mbox{if }\left(\frac{p}{3}\right)=\left(\frac{p}{5}\right)=+1\\
3u^2-5v^2&\mbox{if }\left(\frac{p}{3}\right)=\left(\frac{p}{5}\right)=-1.\end{array}\right.
\label{c5}\end{equation}
Hence Conjecture~1 holds at $n=5$ and the cores of $(c_5^+(p),\,c_5^-(p))$
are $(15,\,1)$ for $p\equiv1\mbox{ or 4 mod 15}$, or $(3,\, 5)$ for $p\equiv2\mbox{ or 8 mod 15}$.
Underlying this great simplicity is a modular form
of weight 3 and level 15, with a Fourier series, in the upper half of the complex $\tau$-plane,
\begin{equation}
f_{3,15}(\tau)=\left(\eta_3\eta_5\right)^3+\left(\eta_1\eta_{15}\right)^3=\sum_{n>0}A_5(n)q^n
\label{f5}\end{equation}
where $q:=\exp(2\pi i\tau)$, $\eta_n:=\eta(q^n)$, $\eta(q):=q^{1/24}\prod_{k>0}(1-q^k)$, $A_5(1)=1$
and, for prime $p$, $A_5(p)=c_5(p)$.

\subsection{A Hecke eigenform at $n=7$}%5.2

At $n=7$, with $y=p^2T$, we have  $Z_7(3,T)=1-10T+y^2$, $Z_7(5,T)=1-y^2$,
$Z_7(7,T)=1-70T+y^2$ and $Z_7(11,T)=(1-y)^2(1+y)$. Thus
$c^\pm_7(3)=2(9\mp5)$, $c^\pm_7(5)=0$, $c^\pm_7(7)=14(7\mp5)$
and $c_7^\pm(11)=0$ all conform to Conjecture~1, since they are non-negative
integers and when non-zero have cores that divide $2\times3\times7=42$.

For $\left(\frac{p}{105}\right)=\pm1$ we have $c_7^\pm(p)=0$. For $p=2$ and for $11<p<10^5$,
$c_7^\mp(p)/2=p^2\pm c_7(p)$ vanishes only at $p=48281$. Otherwise it is a
positive integer whose core divides 42. 
Hence Conjecture~1 holds at $n=7$ for all primes $p<10^5$.

The answer 42, for the least common multiple of cores, comes from 
the existence of a weight-3 newform on $\Gamma_0(525)$, with quartic nebentypus
and Hecke eigenvalues in ${\bf Q}(\sqrt{-1},\sqrt{6},\sqrt{14})$. For prime $p$ coprime to 105,
its $p$-th eigenvalue, $\lambda_p$, determines 
\begin{equation}c_7(p)=\left(\frac{p}{105}\right)\left(|\lambda_p|^2-p^2\right)
\label{c7}\end{equation}
as conjectured~\cite{E} by Evans and Stein and proven~\cite{YV} by Yun and Vincent.

For $\left(\frac{p}{105}\right)=\pm1$, we have $c_7^\mp(p)/2=|\lambda_p|^2$,
which vanishes at $p=11$ and at $p=48281$. If $\lambda_p$ is non-zero, then
the core $\mu$ of $|\lambda_p|^2$ is determined by a signature composed of the signs
$\left(\frac{p}{q}\right)$, for $q=3$, 5 and 7, as follows:
\begin{equation}
\mu(+,+,+)=1,\,\mu(-,-,+)=3,\,\mu(+,-,-)=7,\,\mu(-,+,-)=21,
\label{mup}\end{equation}
\begin{equation}
\mu(-,-,-)=2,\,\mu(+.+,-)=6,\,\mu(-,+,+)=14,\,\mu(+,-,+)=42.
\label{mum}\end{equation}
This map from signatures to cores is multiplicative: if primes
$p_1$, $p_2$ and $p_3$ give cores $\mu_1$, $\mu_2$ and $\mu_3$,
and $\left(\frac{p_3}{q}\right)=\left(\frac{p_1p_2}{q}\right)$, 
for $q=3$, 5 and 7, then $\mu_3$ is the core of $\mu_1\mu_2$.
Hence 3 values of $\mu$, at the primes 2, 13 and 17, with signatures 
$(-,-+)$, $(+,-,-)$ and $(-,-,-)$, determine all 8 cores  in~(\ref{mup},\ref{mum}).

\subsection{Cores at $n=9$ and $n=11$}%5.3

For $n=9$, with $y=p^3T$, 
we have $Z_9(3,T)=1+6T+y^2$, $Z_9(5,T)=(1+y)(1-106T+y^2)$,
and $Z_9(7,T)=(1-y)(1+238T+y^2)$, for $p|105$. In these cases, $c_7^\pm(p)$
either vanishes or is positive, with a core dividing 210, in accord with Conjecture~1.
For $\left(\frac{p}{105}\right)=-1$, we have $Z_9(p,T)=(1-y^2)(1-c_9(p)T+y^2)$
and hence $c_7^\pm(p)=0$. The interesting case is $\left(\frac{p}{105}\right)=1$
for which the irreducible quartic polynomial
\begin{equation}
Z_9(p,T)=1-c_9(p)(T+p^6T^3)+\mbox{$\frac12$}(c^2_9(p)-c_9(p^2))T^2+p^{12}T^4
\label{Z9}\end{equation}
requires computations in both ${\bf F}_p$ and ${\bf F}_{p^2}$. 
At $p=2$ we find that $c_9^+(2)=5\times3^2$
and $c_9^-(2)=21$, while at $p=13$ we have $c_9^+(13)=105\times24^2$
and $c_9^-(13)=80^2$. There is a regular pattern:
\begin{equation}
(c^+_9(p),\,c^-_9(p))=\left\{\begin{array}{cl}
(105u^2,\,v^2)&\mbox{if }\left(\frac{p}{3}\right)=\left(\frac{p}{35}\right)=+1\\
(5u^2,\,21v^2)&\mbox{if }\left(\frac{p}{3}\right)=\left(\frac{p}{35}\right)=-1\end{array}\right.
\label{c9}\end{equation}
with positive integers $(u,\,v)$ determined by $c_9(p)$ and $c_9(p^2)$,
for all $p\leq631$ having the signatures in~(\ref{c9}). Since $c_9(p)=O(p^3)$ and
$c_9(p^2)=O(p^6)$, much economy is achieved by the pair $(u,v)$, with $u=O(v)=O(p^2)$.
In Section~8, I shall show how to achieve even greater economy. 

For $n=11$, with $y=p^4T$, 
we have $Z_{11}(3,T)=(1-y)(1+150T+y^2)$ and $Z_{11}(p,T)$ vanishing at $y=\pm1$,
for $p=5$, 7 and 11. Thus we need compute only $c^-_{11}(3)=6^3$ to check
Conjecture~1 for $p|1155$. For $\left(\frac{p}{1155}\right)=\pm1$, we have
$c^\pm_{11}(p)=0$ and $c^\mp(p)$ an even positive integer for $p\leq631$, with the
exception of the case $p=17$, where 
\begin{equation}
Z_{11}(17,T)=(1-y)(1+y)^2(1-27290T+y^2)
\label{Z11_17}\end{equation}
and hence $c^+_{11}(17)=c^-_{11}(17)=0$. When $c^\mp_{11}(p)/2$ is non-zero,
its core, $\mu$, is determined by the signature
$\left(\frac{p}{q}\right)$, for $q=3$, 5, 7 and 11, according to a multiplicative
map generated by 
$\mu(+,+,-,-)=2$, $\mu(+,-,-,+)=3$, $\mu(+,-,+,+)=5$ and $\mu(+,-,-,-)=11.$
Thus all the cores divide 330.

\subsection{Doctrine of signatures}%5.4

For odd positive integer $n$, let $n!!:=\prod_{0\leq2k<n} (n-2k)$
be the product of all odd positive integers not exceeding $n$.
Let $\omega(n!!)$ be the number of distinct  primes $q|n!!$.
Let the signature of a prime $p$ coprime to $n!!$
be the list of $\omega(n!!)$ signs $\left(\frac{p}{q}\right)$ for primes $q|n!!$.

{\bf Conjecture 2.} For $n\equiv1$ mod 4, $n\ge5$  and $\left(\frac{p}{n!!}\right)=1$,
$(c^+_n(p),\,c^-_n(p))$ is a pair of positive integers whose cores $(\mu^+,\,\mu^-)$
are determined by a multiplicative map, from signatures to pairs of cores.
Moreover, for each pair, the core of $\mu^+\mu^-$ coincides with the core of $n!!$.

{\bf Conjecture 3.} For $n\equiv3$ mod 4, $n\ge7$ and $\left(\frac{p}{n!!}\right)=\pm1$, 
$c^\mp_n(p)/2$ is a non-negative integer. If this integer is non-zero, its core $\mu$
is determined by a multiplicative map, from signatures to cores.

These agree with the previous findings. At $n=5$ the pairs of
cores are $(15,\,1)$ and $(3,\, 5)$; at $n=9$ they are $(105,\, 1)$
and $(5,\,21)$. At $n=7$, the cores are the 8 divisors of 42; at
$n=11$ they are the 16 divisors of 330.

\newpage

\subsection{Tests at $n\ge13$}%5.5

At $n=13$, the primes dividing $13!!$ are 3, 5, 7, 11 and 13,
with product 15015, which is also the core of $13!!$.
At these 5 primes, Conjecture~1 holds, with non-zero values
for $c^+_{13}(3)$ and for all 5 cases of $c^-_{13}(p)$ with $p|15015$.
Conjecture~2 holds for $\left(\frac{p}{15105}\right)=1$ and $p\leq109$,
with a map generated by 
$\mu^-(-,+,-,+,+)=2$, 
$\mu^-(-,-,+,+,+)=5$, 
$\mu^-(-,+,+,-,+)=11$ and  
$\mu^-(+,+,+,-,-)=13$. In all cases the core of $\mu^+\mu^-$ is 
15015. Hence $\mu^-$ is a divisor of
$2\times5\times11\times13=1430$ and $\mu^+$
is a divisor of 30030. 
The map was determined by data at 4 of the primes
with $\left(\frac{p}{15105}\right)=1$ and tested at the remaining 
12 cases with $p\leq109$.

At $n=15$, Conjecture~1 holds at the primes $p|15015$,
where $c^+_{15}(3)$, $c^-_{15}(3)$ and $c^+_{15}(5)$ are non-zero.
Conjecture~2 holds for $\left(\frac{p}{15015}\right)=\pm1$ and $p\leq109$,
with a map generated by 
$\mu(+,+,+,-,+)=3$, 
$\mu(-,-,+,-,+)=5$, 
$\mu(+,+,+,-,-)=7$,  
$\mu(+,-,-,-,-)=11$ and
$\mu(+,+,-,-,+)=13$.
Hence $\mu$ is a divisor of 15015.
The map was determined by data at 5 of the primes
with $\left(\frac{p}{15105}\right)=\pm1$ and tested at  
17 others cases with $p\leq109$.

At $n=17$ and $n=19$, we require data on $c_n(p^k)$ with $k\leq4$
to test the conjectures. Algorithm~2 thus
requires us to access a table of traces more than $2p^8$ times.
This becomes a daunting task at quite modest $p$. 
Tests of the conjectures were conducted for $p\leq31$, at no small cost.

At $n=17$, Conjecture~1 holds for the odd primes up to 17, 
with 7 non-zero values of $c^\pm_{17}(p)$, of which the largest is
$c^-_{17}(17)=715\times135085536^2$. The two primes $p\leq31$
with $\left(\frac{p}{17!!}\right)=1$ give
$(c^+_{17}(2),\,c^-_{17}(2))=(1105\times3^4,\,385\times3^2)$
and $(c^+_{17}(29),\,c^-_{17}(29))=
(35\times172876746240^2,\,12155\times1827678336^2)$.
In both cases the core of $\mu^+\mu^-$ is $17017$,
which is indeed the core $17!!$.

At $n=19$, Conjecture~1 holds for the odd primes up to 19, 
with 11 non-zero values of $c^\pm_{19}(p)$, of which the largest two are
$c^+_{19}(17)=285\times167446803456^2$ and
$c^-_{19}(17)=42\times1239609466368^2$.
Primes $p\leq31$ with $\left(\frac{p}{19!!}\right)=\pm1$ yield
\begin{equation}
\mu(-,-,+,-,-,+,-)=105,\quad
\mu(-,-,+,+,+,-,+)=19\times105,
\label{mu19a}\end{equation}
\begin{equation}
\mu(-,+,+,-,+,-,-)=13\times105,\quad
\mu(+,+,-,+,-,-,-)=7\times19,
\label{mu19b}\end{equation}
with $c^-_{19}(31)/2=7\times19\times3787940863744^2$
obtained by accessing a table of traces in ${\bf F}_{31^4}$ more than
$1.7\times10^{12}$ times, by efficient algorithms, below.

\section{Efficient computation of $c_n(p^k)$ and $Z_n(p,T)$}%6

The trace in the definition
\begin{equation}
K(a):=\sum_{x\in{\bf F}_q^*}\exp\left(\frac{2\pi i}{p}\trace\left(x+\frac{a}{x}\right)\right)
\label{Kagain}\end{equation}
of a Kloosterman sum is a trace of Frobenius in ${\bf F}_q$ over ${\bf F}_p$.
For  $q=p^k$, it maps an element $z$ of ${\bf F}_q$, in this case $z=x+a/x$, to an element
\begin{equation}
\trace(z):=\sum_{j=0}^{k-1}z^{p^j}=z+z^p+z^{p^2}+\ldots+z^{q/p}
\label{trace}\end{equation}
of ${\bf F}_p$, which we may take to be an integer in $[0,p-1]$, modulo $p$, giving a $p$-th root
unity in the summand of (\ref{Kagain}). We need $q-1$ 
traces to evaluate the sum in~(\ref{Kagain}) over all elements
of ${\bf F}_q$ except 0. We need $K(a)$ for  $q-1$ elements of ${\bf F}_q$, 
to evaluate the sum over symmetric powers in
\begin{equation}
c_n(q):=-\frac{1+S_n(q)}{q^2},\quad
S_n(q):=\sum_{a\in{\bf F}_q^*}\sum_{j=0}^n\left[g(a)\right]^j\left[h(a)\right]^{n-j}
\label{cagain}\end{equation}
with $K(a)=-g(a)-h(a)$ and $g(a)h(a)=q$. As written,~(\ref{Kagain},\ref{cagain}) appear
to ask for $(q-1)^2$ traces, each with $k$ terms in (\ref{trace}). With $q=31^4$,  we
do not want to evaluate $4(31^4-1)^2>3\times10^{12}$ powers of elements of ${\bf F}_q$.
The following algorithm shows how to avoid this.

\subsection{Algorithm for $c_n(p^k)$}%6.1

{\bf Algorithm 2.} [Evaluation of  $c_n(q)$, for $q=p^k$ and all $n\in[1,N]$.]
\begin{enumerate}
\item Find a monic polynomial, $f$, of degree $k$, that is irreducible over ${\bf F}_p[x]$.\\{}
[For $q=31^4$, we may use $f\equiv x^4+x^3+2x^2-4x+3\mbox{ mod 31}$.]
\item Find a polynomial, $g$, that is, modulo $f$, a primitive root for ${\bf F}_q^*$.\\{}
[In the example, we may use $g\equiv7x^2 + 9x + 14\mbox{ mod }f$, since 
$g^{(q-1)/d}-1$ is non-zero for all divisors $d|(q-1)$ with $d>1$.]
\item For $m\in[0,p-1]$, store in $C[m]$ the value of $\cos(2\pi m/p)$, at suitable
numerical precision. [The sum in~(\ref{Kagain}) is real, so cosines suffice.]
\item For $n\in[1,N]$ and $m\in[1,N]$, store in $U[n,m]$ the integer coefficient of $x^n y^m$
in the expansion of $1/(1+xy+x^2q)$. [This vectorizes~(\ref{cagain}).]
\item  For $m\in[0,k-1]$, store in $X[m]$ the trace of $z^m$, with $z\equiv x\mbox{ mod }f(x)$.
[In the example, we obtain, from~(\ref{trace}), $X\equiv[4, 30, 28, 17]\mbox{ mod } 31$.]
\item For $r\in[1,q-1]$, store in $T[r]$ the trace of $g^r$, using the data in $X$, as follows. 
Set $r=0$, $t=1$. While $r<q-1$, add 1 to $r$, multiply $t$ by $g$, 
set $T[r]=\sum_m t_m X[m]$, where $t_m$ is the coefficient of $x^m$ in $t$.
[In the example, $g$ gives $T[1]=7X[2]+9X[1]+14X[0]\equiv26\mbox{ mod 31}$, then
$g^2=15x^3 + 24x^2 + 14x + 18$ gives $T[2]\equiv28\mbox{ mod 31}$, and
so on.]
\item For $a\in[1,q-1]$, store $K[a]=\sum_{0<r<q} C[m(a,r)]$, with $m(a,r)\in[0,p-1]$ 
given by $m(a,r)\equiv T[r]+T[s(a,r)]\mbox{ mod }p$, where $s(a,r)\in[1,q-1]$ is given by 
$s(a,r)\equiv (a-r)\mbox{ mod }(q-1)$. 
\item For $m\in[1,N]$, store $V[m]=\sum_{0<a<q}(K[a])^m$.
\item For $n\in[1,N]$, compute $S[n]=\sum_{0<m\leq N}U[n,m]V[m]$ and
return $c_n(q)$ as the integer nearest to $-(1+S[n])/q^2$.
\end{enumerate}
For large $q$, most time is spent in Step~7, where $T$ is accessed
$2(q-1)^2$ times. The entries in $T$ are determined in Step~6 by
$(q-1)$ multiplications in ${\bf F}_q$ and $O(kq)$ accesses to $X$. The $k$
entries in $X$ are determined in Step~5 by binary exponentiations
requiring $O(k^2\log(q)/\log(2))$ multiplications.

\subsection{Algorithm for $Z_n(p,T)$}%6.2

{\bf Algorithm 3.} [Determination of $Z_n(p,T)$ for all $n\in[1,N]$.]
\begin{enumerate}
\item Use Theorem 1 and Algorithm 1 to determine the largest $d(p,n)$
for $n\in[1,N]$ and record this as $D$.
If $D>0$, then use Algorithm~2, for $k\in[1,D]$, and store
$C[n,k]=c_n(p^k)$, for $n\in[1,N]$.
\item For $n\in[1,N]$, 
set $d=d(p,n)$, $y=p^{\lfloor(n-3)/2\rfloor}T$, $z=p^{n-3}T^2$.
Find $F(T)=(1-y)^{a(p,n)}(1+y)^{b(p,n)}$,
using (\ref{abodd}), or (\ref{abgen},\ref{bgen}), or (\ref{abodd2}), or (\ref{aeven2},\ref{beven2}),
according as the parities of $n$ and $p$. Develop the expansion
$\exp(-\sum_{0<k\leq d}C[n,k]T^k/k)/F(T)=\sum_{0\leq k\leq d}m_kT^k+O(T^{d+1})$.
Return $Z_n(p,T)=F(T)(m_dT^d+\sum_{0\leq k<d}m_k(1+z^{d-k})T^k)$.
\end{enumerate}
This minimizes the number, $D$, of uses of Algorithm~2 and maximally exploits
the functional equation~(\ref{Msym}). To obtain only $Z_{2k}(p,T)$, set
$D$ to the largest $d(p,2k)$ for $2k\leq N$, which is often smaller than the $D$ above. 

\subsection{Features of the database}%6.3

The database compiled with these algorithms contains
determinations of $Z_n(p,T)$ for 
$p<10^5$ at $n\leq8$; for $p\leq631$ at $n\leq12$; 
for $p\leq109$ at $n\leq16$; for $p\leq31$ at $n\leq20$. 

At small fixed $p$, Algorithm~3 was used to extend the range of $n$ as follows. 
At $p=2$, determinations were made for even $n\leq200$ and odd $n\leq65$;
at $p=3$ for even $n\leq116$ and odd $n\leq59$; at $p=5$ for even $n\leq44$
and odd $n\leq35$; at $p=7$ and $p=11$ for all $n\leq27$.

The results are available in files, on serious request to the author. They conform
with Conjectures~1, 2 and 3, at all listed $p$ and odd $n$. 

For even $n$, $M_n(p,T)$ was always found to be irreducible, with $d(p,n)$ pairs of
complex roots at  $T=\exp(\pm i\theta)p^{{(3-n)}/2}$ and $0<\theta<\pi$.
For odd $n$, only $M_7(11,T)=(1-y)^2$, $M_7(48281,T)=(1-y)^2$ and 
$M_{11}(17,T)=(1+y)^2(1-27290T+y^2)$ were found to be reducible.

For odd $n<p$, the choice of exponent in~(\ref{mpm}) is almost always minimal.
There is only one case in the database with non-zero $c^\pm_n(p)$ divisible by $p$,
for odd $n<p$, namely at $n=13$ and $p=109$, where
\begin{eqnarray}
c^+_{13}(109)&=&3\times5\times7\times11\times109^2\times515328^2\label{c109p}\\
c^-_{13}(109)&=&13\times109^2\times2399960^2\label{c109m}
\label{c109}\end{eqnarray}
are in full accord with Conjectures~1 and 2, but also divisible by $p^2$.

From Corollary~1, we obtain a bound  $d(2,2k)=d(3,2k)\leq\left\lfloor\frac{k+1}{6}\right\rfloor$.
This relatively slow growth enables one to determine
very large values of $c_n(q)$ when $n$ is even and $q$ has characteristic
2 or 3. Here are examples yielding probable primes, of sizes ranging from
3238 to 4366 decimal digits:
\begin{eqnarray}
p_{3238}=-2^{-157}c_{142}(2^{157}),&&
p_{3449}=2^{-632}c_{156}(2^{158}),\label{bp1}\\
p_{3614}=2^{-2}3^{-279}c_{116}(3^{139}),&&
p_{3638}=-c_{82}(3^{193}),\label{bp2}\\
p_{3903}=2^{-174}c_{154}(2^{174}),&&
p_{4366}=-2^{-200}c_{150}(2^{200}).\label{bp3}
\end{eqnarray}

\section{L-series and Feynman integrals}%7

For $n<8$ and $s$ with a suitably large real part, I define the L-series
\begin{equation}
L_n(s):=\prod_{p\geq2}\frac{1}{Z_n(p,p^{-s})}=\sum_{m>0}\frac{A_n(m)}{m^s}
\label{Ln}\end{equation}
with $A_n(1)=1$, $A_n(m_1m_2)=A_n(m_1)A_n(m_2)$, for $\gcd(m_1,m_2)=1$, and $A_n(p)=c_n(p)$
at prime $p$. Then $L_1(s)=L_2(s)=1$ and from~(\ref{Z34}) we obtain
\begin{eqnarray}
L_3(s)&=&\sum_{m>0}\left(\frac{m}{3}\right){m^{-s}}=
\sum_{k\geq0}\left(\frac{1}{(3k+1)^s}-\frac{1}{(3k+2)^s}\right)\label{Z3}\\
L_4(s)&=&(1-2^{-s})\zeta(s)=\sum_{k\geq0}\frac{1}{(2k+1)^s}\label{Z4}
\end{eqnarray}
with a functional equation 
\begin{equation}
\Lambda_3(s):=\left(\frac{3}{\pi}\right)^{s/2}\Gamma\left(\frac{s+1}{2}\right)L_3(s)=\Lambda_3(1-s)
\end{equation}
for the Dirichlet L-series~(\ref{Z3}), giving analytic continuation to $L_3(0)=\sqrt3L_3(1)/\pi=\frac13$,
while at $n=4$ Riemann and Euler give us $L_4(0)=0$ and $L_4(2)=\pi^2/8$. 

Now consider Feynman diagrams, like those illustrated below,
evaluated, in two spacetime dimensions,  by integrals of the form
\begin{equation}
S_{n,s}:=2^s\int_0^\infty[I_0(t)]^{n-s-1}[K_0(t)]^{s+1}t\,dt
\label{Sns}\end{equation}
with $n$ Bessel functions
and loop-number (i.e.\ first Betti number) $s$ satisfying
$s<n\leq2s+2$ and $s>1$ for $n=2s+2$, to ensure convergence.

The internal scalar particles have unit mass
and account for the Bessel function $K_0(t)$ in the integrand. Hence
the two-loop vacuum integral $S_{3,2}$ 
has a propagator $K_0(t)$ associated with each of its three internal edges.
$S_{4,2}:=2^2\int_0^\infty I_0(t)K_0^3(t)\,t\,dt$ is a 
two-loop on-shell sunrise diagram,
with the Bessel function $I_0(t)$ coming from external half-edges,
whose momenta are on the unit mass shell. The one-loop diagram $S_{3,1}$
is obtained from $S_{3,2}$ by cutting an internal edge. Removing the external half-edges
from $S_{3,1}$, we obtain the one-loop vacuum diagram $S_{2,1}$. If we join up the
half-edges in $S_{4,2}$, we obtain a three-loop vacuum diagram, 
$S_{4,3}$. 

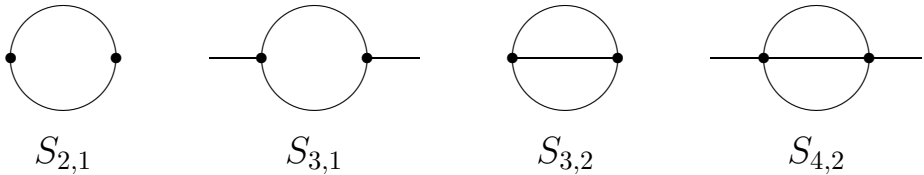
\mbox{\hspace{1cm}}\hfill
\dia{
\put(0,0){\circle{100}}
\put(50,0){\vtx}
\put(-50,0){\vtx}
}{$S_{2,1}$}
\dia{
\put(-100,0){\line(1,0){50}}
\put(50,0){\line(1,0){50}}
\put(0,0){\circle{100}}
\put(50,0){\vtx}
\put(-50,0){\vtx}
}{$S_{3,1}$}
\dia{
\put(-50,0){\line(1,0){100}}
\put(0,0){\circle{100}}
\put(50,0){\vtx}
\put(-50,0){\vtx}
}{$S_{3,2}$}
\dia{
\put(-100,0){\line(1,0){200}}
\put(0,0){\circle{100}}
\put(50,0){\vtx}
\put(-50,0){\vtx}
}{$S_{4,2}$}
\mbox{\hspace{1cm}}\par

\newpage

For $n\leq4$, the $s$-loop
integral $S_{n,s}$ is an {\em integer} multiple~\cite{BBBG} of $L_n(s)$:
\begin{eqnarray}
S_{1,0}&=&L_1(0)=1\label{S10}\\
S_{2,1}&=&L_2(1)=1\label{S21}\\
S_{3,1}&=&2L_3(1)=\frac{2\pi}{\sqrt{3^3}}\label{S31}\\
S_{3,2}&=&3L_3(2)=3\sum_{k\geq0}\left(\frac{1}{(3k+1)^2}-\frac{1}{(3k+2)^2}\right)\label{S32}\\
S_{4,2}&=&2L_4(2)=\frac{\pi^2}{4}\label{S42}\\
S_{4,3}&=&8L_4(3)=7\zeta(3)\label{S43}.
\end{eqnarray}

\subsection{Proofs for 5 Bessel functions}%7.1

In a conference talk,
{\em Reciprocal PSLQ and the tiny nome of Bologna}, given in June 2007
at the Zentrum f\"{u}r interdisziplin\"{a}re Forschung in Bielefeld,
I presented the empirical evaluation~\cite{Bnome}
\begin{equation}
C:=\frac{S_{5,3}}{8\pi^2} =
\frac{\pi}{16}\left(1-\frac{1}{\sqrt{5}}\right)
\left(\sum_{n=-\infty}^\infty e^{-n^2\pi\sqrt{15}}\right)^4
\label{nome}\end{equation}
for the on-shell 3-loop sunrise diagram $S_{5,3}$ in two spacetime dimensions.
This implies a neat evaluation as a product of values of the gamma function~\cite{Lap}
\begin{equation}
S_{5,3} =
\frac{1}{30\sqrt{5}}\prod_{k=0}^3\Gamma\left(\frac{2^k}{15}\right)
\label{gamma}\end{equation}
by applying the Chowla-Selberg theorem to elliptic integrals at the 15th singular value~\cite{BBBG}.
Intense work in 2007 with Jon Borwein, in Halifax, Nova Scotia, showed that~(\ref{gamma})
is equivalent to
\begin{equation}
S_{5,3} = \frac{4\pi} {\sqrt{15}}\,S_{5,2}.
\label{S5rel}\end{equation}
Discussions with Spencer Bloch and Francis Brown,
at a summer school in Les Houches, organized by Dirk Kreimer in 2010, pointed to a
connection with the modular form $f_{3,15}=(\eta_3\eta_5)^3+(\eta_1\eta_{15})^3$
of Section~5.1. This came from the representation of
\begin{equation}
S_{5,3}=\int_0^\infty\int_0^\infty\int_0^\infty\frac{da\,db\,dc}{
(abc+ab+bc+ca)(a+b+c)+ab+bc+ca}
\label{schw}\end{equation}
as an integral over Schwinger parameters. Then counts of the zeros of the denominator
of the integrand, in finite fields ${\bf F}_p$, implicated the Fourier coefficients $A_5(p)=c_5(p)$
of $f_{3,15}$, at small primes. 

Thus I was led to investigate the L-series obtained by setting $n=5$ in~(\ref{Ln}).
This has a functional equation
\begin{equation}
\Lambda_5(s):=\left(\frac{15}{\pi^2}\right)^{s/2}
\Gamma\left(\frac{s}{2}\right)\Gamma\left(\frac{s+1}{2}\right)L_5(s)=\Lambda_5(3-s)
\label{Lam5}\end{equation}
and analytic continuation yields a convergent expansion for
\begin{equation}
\frac{2\pi}{\sqrt{15}}L_5(1)=L_5(2)=\sum_{n>0}\frac{A_5(n)}{n^2}
\left(1+\frac{4\pi n}{\sqrt{15}}\right)\exp\left(-\frac{2\pi n}{\sqrt{15}}\right).
\label{L5crit}\end{equation}
Numerical comparison with~(\ref{gamma}) then gave the evaluations
\begin{eqnarray}
S_{5,2}&=&3L_5(2)\label{S52}\\
S_{5,3}&=&\frac{48}{5}\zeta(2)L_5(1).\label{S53}
\end{eqnarray}
The neat result~(\ref{S52}) was proven by combining
my work with Bailey, Borwein and Glasser~\cite{BBBG}, on $S_{5,2}$, with work by
Rogers, Wan and Zucker~\cite{RWZ} on $L_5(2)$. The proof of~(\ref{S53})
follows from work by Bloch, Kerr and Vanhove~\cite{BBV} on $S_{5,3}$.
These authors evaluated the 3-loop sunrise diagram off the external mass shell, in terms
of an elliptic trilogarithm. Delicate work on the on-shell limit, elucidated by Detchat Samart~\cite{S},
then gave a reduction to gamma values. Hence all the equations~(\ref{nome})
to~(\ref{S53}) now have proofs, after 8 years of hard work.

\subsection{Conjectures for 5 Bessel functions}%7.2

The 4-loop vacuum integral with 5 Bessel functions has a representation
\begin{equation}
S_{5,4}=2\int_0^\infty\int_0^\infty\int_0^\infty\frac{\log(a+b+c+1)\,da\,db\,dc}
{(abc+ab+bc+ca)(a+b+c)+ab+bc+ca}
\label{schwv}\end{equation}
with the same denominator as for $S_{5,3}$ in~(\ref{schw}), but a logarithmic numerator,
resulting from integration over an extra Schwinger parameter. This resisted 8 years of effort
to find a relation to the L-series $L_5(s)$ until Anton Mellit and I met at the Mainz Institute
for Theoretical Physics in 2015 and experimented with determinants of 
matrices of Bessel moments.
On the basis of numerical investigation, we arrived at the conjectures
\begin{eqnarray}
\det\int_0^\infty I_0(t)K_0^3(t)\left [\begin{array}{lr}
K_0(t)&t^2K_0(t)\\
I_0(t)&t^2I_0(t)\end{array}\right]t\,dt
&\stackrel{?}{=}&\frac{2\pi^3}{\sqrt{3^35^3}}\label{con5i}\\
\det\int_0^\infty K_0^3(t)\left [\begin{array}{lr}
K_0^2(t)&t^2K_0^2(t)\\
I_0^2(t)&t^2I_0^2(t)\end{array}\right]t\,dt
&\stackrel{?}{=}&\frac{45}{8\pi^2}L_5(4)\label{con5k}
\end{eqnarray}
with question marks indicating that these evaluations are as yet unproven.
 
Our rationale for these constructions was as follows. 
In 2007, reciprocal PSLQ for the matrix
whose determinant is taken in~(\ref{con5i}) gave
\begin{equation}
\left[\begin{array}{lr}
\pi^2C&\pi^2\left(\frac{2}{15}\right)^2\left(13C-\frac{1}{10C}\right)\\
\frac{\sqrt{15}\pi}{2}C&\frac{\sqrt{15}\pi}{2}\left(\frac{2}{15}\right)^2\left(13C+\frac{1}{10C}\right)
\end{array}\right]
\label{mcon5i}\end{equation}
and since then the first column and second row have been proven. The entry in row 1 and column 2
is conjectural, though checked to 1000-digit precision. It has precisely the form
to make the determinant independent of $C$.
In~(\ref{con5k}),
we chose a matrix whose second row is identical to that in~(\ref{con5i}), but with vacuum
integrals appearing in the first row, and were rewarded by a result for $L_5(4)$,
outside the critical strip. Conjectures~(\ref{con5i},\ref{con5k})
may be combined with proven results to obtain the striking evaluation
\begin{eqnarray}
\frac{L_5(4)}{L_5(2)\zeta(2)}&\stackrel{?}{=}&\frac{4}{5}\int_0^\infty
(R-t^2)K_0^5(t)\,t\,dt \label{L54}\\
R&:=&13\left(\frac{2}{15}\right)^2+32\,\prod_{k=0}^3\frac{\Gamma(1-2^k/15)}{\Gamma(2^k/15)}\label{KK}
\end{eqnarray}
with our lucky integer 13 inferred from~(\ref{mcon5i}). It would have been hard to arrive
at~(\ref{L54},\ref{KK}) without taking determinants in~(\ref{con5i},\ref{con5k}).

\subsection{Conjectures for 6 Bessel functions}%7.3

At $n=6$, I found, with help from Francis Brown at Les Houches in 2010, a
modular form of weight 4 and level 6 
\begin{equation}
f_{4,6}(\tau)=\left(\eta_1\eta_2\eta_3\eta_6\right)^2=\sum_{n>0}A_6(n)q^n
\label{f6}\end{equation}
with $A_6(p)$ at the primes agreeing with counts in ${\bf F}_p$
of zeros of the denominator of the Feynman integrand for the 4-loop
sunrise diagram $S_{6,4}$, with 6 Bessel functions. 
Work by Hulek, Spandaw, van Geemen and van Straten~\cite{HSGS}  showed
that~(\ref{f6}) solves the Kloosterman problem at $n=6$, giving
$A_6(p)=c_6(p)$ at the primes.
Then the functional equation
\begin{equation}
\Lambda_6(s):=\left(\frac{6}{\pi^2}\right)^{s/2}
\Gamma\left(\frac{s}{2}\right)\Gamma\left(\frac{s+1}{2}\right)L_6(s)=\Lambda_6(4-s)
\label{Lam6}\end{equation}
yielded convergent  expansions for
\begin{eqnarray}
L_6(2)&=&\sum_{n>0}\frac{A_6(n)}{n^2}
\left(2+\frac{4\pi n}{\sqrt{6}}\right)\exp\left(-\frac{2\pi n}{\sqrt{6}}\right)\label{L6crit2}\\
2\zeta(2)L_6(1)=L_6(3)&=&\sum_{n>0}\frac{A_6(n)}{n^3}
\left(1+\frac{2\pi n}{\sqrt{6}}+\frac{2\pi^2n^2}{3}\right)\exp\left(-\frac{2\pi n}{\sqrt{6}}\right)
\quad{}\quad{}\label{L6crit13}
\end{eqnarray} 
at the three integers inside the critical strip, with $0<s<4$, and numerical investigation
led to the conjectures
\begin{eqnarray}
S_{6,2}&\stackrel{?}{=}&6L_6(2)\label{S62}\\
S_{6,3}&\stackrel{?}{=}&12L_6(3)=24\zeta(2)L_6(1)\label{S63}\\
S_{6,4}&\stackrel{?}{=}&48\zeta(2)L_6(2)\label{S64}
\end{eqnarray}
which have been checked at 1000-digit precision. It is notable that
$S_{6,4}$ is pulled down to $L_6(2)$, by a multiple of $\zeta(2)$. 
Hence conjectures~(\ref{S62},\ref{S64}) imply the sum rule~\cite{BBBG}
\begin{equation}
\int_0^\infty I_0(t)K_0^3(t)\left(\pi^2I_0^2(t)-3K_0^2(t)\right)t\,dt\;\stackrel{?}{=}\;0.
\label{sum6}\end{equation}

It was harder to relate Feynman integrals 
to $L_6(5)$, outside the critical strip. This problem was cracked
by using determinants in
conjectures
\begin{eqnarray}
\det\int_0^\infty I_0(t)K_0^4(t)\left [\begin{array}{lr}
K_0(t)&t^2K_0(t)\\
I_0(t)&t^2I_0(t)\end{array}\right]t\,dt
&\stackrel{?}{=}&\frac{5}{32}\zeta(4)\label{con6i}\\
\det\int_0^\infty K_0^4(t)\left [\begin{array}{lr}
K_0^2(t)&t^2K_0^2(t)\\
I_0^2(t)&t^2I_0^2(t)\end{array}\right]t\,dt
&\stackrel{?}{=}&\frac{27}{4\pi^2}L_6(5)\label{con6k}
\end{eqnarray}
that neatly follow the pattern discovered at $n=5$, in~(\ref{con5i},\ref{con5k}).

\subsection{Mahler measures and vacuum diagrams}%7.4

A Laurent polynomial $P(x_1,\ldots, x_n)$ has a logarithmic Mahler measure
\begin{equation}
m(P):=\int_0^1{\rm d}t_1\ldots\int_0^1{\rm d}t_n
\log\left(\left|P\left(e^{2\pi i t_1},\ldots,e^{2\pi i t_n}\right)\right|\right)
\label{mahl}\end{equation}
that may sometimes evaluate to an L-series~\cite{RZ1}. For example
Christopher Deninger~\cite{D} conjectured that 
\begin{equation}
m\left(1+x_1+\frac{1}{x_1}+x_2+\frac{1}{x_2}\right)=\frac{15}{4\pi^2}L_{2,15}(2)
\label{mahl1}\end{equation}
where the L-series comes from the modular form 
$\eta_1\eta_3\eta_5\eta_{15},$ with weight 2 and level 15.
This was proven by Mathew Rogers and Wadim Zudilin~\cite{RZ2}.
David Boyd conjectured~\cite{Boyd} and Anton Mellit proved~\cite{M} that
\begin{equation}
m\left(1+x_1+\frac{1}{x_1}+x_2+\frac{1}{x_2}+x_1x_2+\frac{1}{x_1x_2}\right)=\frac{7}{2\pi^2}L_{2,14}(2)
\label{mahl2}\end{equation}
where the L-series comes from $\eta_1\eta_2\eta_7\eta_{14},$
with weight 2 and level 14.

It is instructive to note that 
\begin{eqnarray}
m(1+x_1+x_2)&=&\frac{\sqrt3}{4\pi}S_{3,2}=\frac{\sqrt{3^3}}{4\pi}L_3(2)\label{m32}\\
m(1+x_1+x_2+x_3)&=&\frac{1}{2\pi^2}S_{4,3}=\frac{7}{2\pi^2}\zeta(3)=\frac{4}{\pi^2}L_4(3)\label{m43}
\end{eqnarray}
evaluate in terms of vacuum diagrams $S_{n,n-1}:=2^{n-1}\int_0^\infty K^n_0(t)\,t\, dt$,
with $n$ Bessel functions at $n-1$ loops.
One might therefore expect the L-series for the 5 and 6-Bessel modular forms,
$(\eta_3\eta_5)^3+(\eta_1\eta_{15})^3$ and $(\eta_1\eta_2\eta_3\eta_6)^2,$
to determine Mahler measures, outside their critical strips, where ~(\ref{con5k},\ref{con6k}) 
give conjectural evaluations of $L_5(4)$ and $L_6(5)$ in terms of determinants
that include $S_{5,4}$ and $S_{6,5}$, respectively. Indeed they do, via relations
\begin{eqnarray}
m(1+x_1+x_2+x_3+x_4)&\stackrel{?}{=}&6\left(\frac{\sqrt{15}}{2\pi}\right)^5L_5(4)\label{m54}\\
m(1+x_1+x_2+x_3+x_4+x_5)&\stackrel{?}{=}&3\left(\frac{\sqrt{6}}{\pi}\right)^6L_6(5)\label{m65}
\end{eqnarray}
conjectured  by Rodriguez Villegas~\cite{V}, listed in~\cite{F} and tested in~\cite{BB}, at 1000-digit
precision, using a Bessel formula~\cite{BSWZ}
\begin{equation}
m(1+x_1+\ldots+x_{n-1})=-\log(2)-\gamma-\int_0^\infty\frac{d J_0^n(t)}{dt}\log(t)\,dt
\label{mahln}\end{equation}
that I had derived from Kluyver's work in 1905 on $n$-step walks~\cite{Kl}. 
Here $\gamma$ is Euler's constant
and the oscillatory Bessel function $J_0(t):=I_0(i t)$
leads to a demanding quadrature, at high precision,
as in the case of off-shell sunrise diagrams with spacelike external data, $w^2<0$.
As far as I know, no-one has related logarithmic Mahler measures to non-critical L-series
from modular forms with weight greater than 4. Nevertheless, the story of
the relation between vacuum diagrams and non-critical L-series, is not finished, as I shall show
at $n=8$, with a modular form of weight 6.

\subsection{Conjectures for 7 Bessel functions}%7.5

Until last year, no relation between L-series and Feynman integrals with 7 Bessel functions
had been discovered. Undaunted by previous failure, Anton Mellit proposed that we use 
my determinations of local factors, from Kloosterman sums in ${\bf F}_q$, in the L-series
\begin{equation}
L_7(s):=\prod_{p\ge2}\frac{1}{Z_7(p,p^{-s})}=\sum_{n>0}\frac{A_7(n)}{n^s}
\label{L7s}\end{equation}
and try to discover a functional
relation that would allow analytic continuation to critical values with $0<s<5$.
Crucial in this endeavour were my determinations of the local factors
\begin{eqnarray}
Z_7(2,2^{-s})&=&\left(1-\frac{1}{2^{s-2}}\right)\left(1+\frac{5}{2^{s-2}}+\frac{1}{2^{2s-4}}\right)\label{Z7s2}\\
Z_7(3,3^{-s})&=&1-\frac{10}{3^{s-2}}+\frac{1}{3^{2s-4}}\label{Z7s3}\\
Z_7(5,5^{-s})&=&1-\frac{1}{5^{2s-4}}\label{Z7s5}\\
Z_7(7,7^{-s})&=&1-\frac{10}{7^{s-3}}+\frac{1}{7^{2s-4}}\label{Z7s7}
\end{eqnarray}
at $p\leq7$, where Evans had been silent~\cite{E}. Thereafter,
it was sufficient to use {\tt Sage}, with commands kindly provided by William Stein,
to determine local factors from Hecke eigenvalues,  
$\lambda_p\in{\bf Q}(\sqrt{-1},\sqrt{6},\sqrt{14})$, at prime $p$,
of the newform on $\Gamma_0(525)$,
with weight 3 and quartic nebentypus, that gives $|\lambda_p|^2=p^2\pm A_7(p)$,
for $\left(\frac{p}{105}\right)=\pm1$. Using computer power kindly provided at 
the Humboldt Universit\"{a}t zu Berlin, I ran {\tt Sage} for several days
and verified that for $7<p<10^5$ the cores of $|\lambda_p|^2$ are divisors of 42, 
determined from signatures by the multiplicative map~(\ref{mup},\ref{mum}), 
except in the cases $p=11$ and $p=48281$, with signature $(-,+,+)$ and $A_7(p)=p^2$,
where $\lambda_p$ vanishes.  

Having determined $A_7(n)$ for $n\leq100000$,
I used Tim Dokchitser's code~\cite{Dok} {\tt computel}, in {\tt Pari-GP}, to determine the viability
of a functional equation
\begin{equation}
\Lambda_7(s):=\left(\frac{105}{\pi^3}\right)^{s/2}\Gamma\left(\frac{s-1}{2}\right)
\Gamma\left(\frac{s}{2}\right)\Gamma\left(\frac{s+1}{2}\right)L_7(s)
\;\stackrel{?}{=}\;\Lambda_7(5-s)
\label{Lam7}\end{equation}
with the crucial factor $\Gamma((s-1)/2)$ found empirically. Then {\tt computel}
professed itself ready to compute $L_7(s)$ and was used at low precision to
determine a conjectural integer 20 in
\begin{equation}
S_{7,4}:=2^4\int_0^\infty I_0^2(t)K_0^5(t)\,t\,dt\;\stackrel{?}{=}20\zeta(2)L_7(2)
\label{S74}\end{equation}
which was then confirmed to 1000-digit precision, in less than 7 hours.

The 4-loop Feynman integral $S_{7,4}$ is the sole integral with 7 Bessel functions
that I have been able to relate to $L_7(s)$. It is not hard to see why there is only one.
The factor $\Gamma((s-1)/2)$ in the functional equation~(\ref{Lam7}) 
seems to render $s=1$ and hence $s=5-1=4$
inaccessible. Inside the critical strip, with $0<s<5$, that leaves only $s=2$, which
is equivalent to $s=5-2=3$, by the functional equation. 

No determinant  was found to permit an excursion to $s=6$, outside the critical strip. 
However a $3\times3$ matrix of moments of 7 Bessel functions,
\begin{equation} 
M_3:=\int_0^\infty I_0(t)K_0^4(t)\left[\begin{array}{rrr}
K_0^2(t)&t^2K_0^2(t)&t^4K_0^2(t)\\
I_0(t)K_0(t)&t^2I_0(t)K_0(t)&t^4I_0(t)K_0(t)\\
I_0^2(t)&t^2I_0^2(t)&t^4I_0^2(t)\end{array}\right]t\,dt
\label{M3}\end{equation}
gave the intriguing numerical result
\begin{equation}
\det M_3\,\stackrel{?}{=}\,\frac{2^4\pi^6}{\sqrt{3^35^57^7}}
\label{detm3}\end{equation}
with the square root of $3^35^57^7$ resonating with the square root of 
$3^35^5$ in~(\ref{con5i}). Defining $M_k$ to be the $k\times k$
matrix with elements
\begin{equation}
(M_k)_{a,b}:=\int_0^\infty[I_0(t)]^a[K_0(t)]^{2k+1-a}t^{2b-1}dt
\label{Mk}\end{equation}
that are moments of $n=2k+1$ Bessel functions, I found
a similar pattern, working up to a $15\times15$ matrix at $n=31$, where
the striking evaluation
\begin{equation}
\det M_{15}\;\stackrel{?}{=}\;\frac{2^{182}\pi^{120}}
{3^{33}\,5^{20}\,7^5\sqrt{11^3\,13^9\,17^{17}\,19^{19}\,23^{23}\,29^{29}\,31^{31}}}
\label{det15}\end{equation}
was found and checked at 500-digit precision.

{\bf Conjecture 4.} The $k\times k$ matrix with elements~(\ref{Mk}) has determinant
\begin{equation}
\det M_k=\prod_{j=1}^k\frac{(2j)^{k-j}\pi^j}{\sqrt{(2j+1)^{2j+1}}}.
\end{equation}
 
\subsection{Conjectures for 8 Bessel functions}%7.6

At $n=8$, we need to modify the definition~(\ref{Ln}), which served well for $n<8$.
I discovered that the modular form
\begin{equation}
f_{6,6}(\tau)=
\left(\frac{\eta_2^3\eta_3^3}{\eta_1\eta_6}\right)^3
+\left(\frac{\eta_1^3\eta_6^3}{\eta_2\eta_3}\right)^3
=\sum_{n>0}A_8(n)q^n
\label{f8}\end{equation}
with weight 6 and level 6, gives $A_8(p)\equiv c_8(p)$ mod $p$, at the primes.
However, we do not have equality between $A_n(p)$ and $c_n(p)$ for 
$n=8$ and prime $p>2$. Instead
I found that
\begin{equation}
c_8(p)=\left\{\begin{array}{rl}A_8(p)&\mbox{if }p=2\\
p^4+A_8(p)&\mbox{if }p>2\end{array}\right.
\label{A8}\end{equation}
and hence, from~(\ref{Z34}), that
\begin{equation}
L_8(s):=\prod_{p\ge2}\frac{Z_4(p,p^{4-s})}{Z_8(p,p^{-s})}=\sum_{n>0}\frac{A_8(n)}{n^s}
\label{L8}\end{equation}
is the L-series corresponding to the modular form~(\ref{f8}), as was recently
proven by Yun~\cite{YV}. Then the functional equation
\begin{equation}
\Lambda_8(s):=\left(\frac{6}{\pi^2}\right)^{s/2}\Gamma\left(\frac{s}{2}\right)
\Gamma\left(\frac{s+1}{2}\right)L_8(s)=\Lambda_8(6-s)
\end{equation}
enables analytic continuation inside the critical strip, $0<s<6$, where
\begin{eqnarray}
L_8(3)&=&\sum_{n>0}\frac{A_8(n)}{n^3}\left(2+\frac{4\pi n}{\sqrt6}
+\frac{2\pi^2n^2}{3}
\right)\exp\left(-\frac{2\pi n}{\sqrt6}\right)\label{L83}\\
L_8(4)&=&\sum_{n>0}\frac{A_8(n)}{n^4}\left(1+\frac{2\pi n}{\sqrt6}
+\frac{4\pi^2n^2}{9}+\frac{4\pi^3n^3}{9\sqrt6}
\right)\exp\left(-\frac{2\pi n}{\sqrt6}\right)\quad{}\label{L84}
\end{eqnarray}
are given by rapidly convergent series. At $s=5$ we obtain
\begin{equation}
L_8(5)=\sum_{n>0}\frac{A_8(n)}{n^5}\left(1+\frac{2\pi n}{\sqrt6}
+\frac{\pi^2n^2}{3}+\frac{2\pi^3n^3}{9\sqrt6}+\frac{\pi^4n^4}{27}
\right)\exp\left(-\frac{2\pi n}{\sqrt6}\right)
\label{L85}\end{equation}
which should be a rational multiple of $\zeta(2)L_8(3)$,
according to experts~\cite{GZ} in Eichler-Shimura-Manin theory. I verified, at 1000-digit
precision, that
\begin{equation}
\frac{L_8(5)}{\zeta(2)L_8(3)}\;\stackrel{?}{=}\;\frac{4}{7}
\label{L85rel}\end{equation}
and expect equality to be soon proven, by at least one expert.

Evaluating~(\ref{L83},\ref{L84}) numerically and using~(\ref{L85rel}), I found that
\begin{eqnarray}
S_{8,3}&\stackrel{?}{=}&8L_8(3)\label{S83}\\
S_{8,4}&\stackrel{?}{=}&36L_8(4)\label{S84}\\
S_{8,5}&\stackrel{?}{=}&216L_8(5)\label{S85}\\
S_{8,6}&\stackrel{?}{=}&864\zeta(2)L_8(4)\label{S86}
\end{eqnarray}
and have checked these evaluations at 1000-digit precision.
Since these four Feynman integrals have evaluations in terms of only
two independent critical values of $L_8(s)$, we obtain two conjectural sum rules,
namely
\begin{eqnarray}
\int_0^\infty\left(9\pi^2I_0^2(t)-14K_0^2(t)\right)I_0^2(t)K_0^4(t)\,t\,dt&\stackrel{?}{=}&0\label{sr1}\\
\int_0^\infty\left(\pi^2I_0^2(t)-K_0^2(t)\right)I_0(t)K_0^5(t)\,t\,dt&\stackrel{?}{=}&0\label{sr2}
\end{eqnarray}
of which the second is merely the tip of a remarkable empirical iceberg.

\subsection{A conjecture dedicated to Richard Crandall}%7.7

Richard Crandall (1947--2012) was proudly part Cherokee, a physicist and a computational number theorist.
He studied under Richard Feynman at Caltech, under Viki Weisskopf at MIT, and worked in the physics
department at Reed College, Oregon, from 1978 until his untimely death from
acute myeloid leukemia. With Carl Pomerance, he wrote a particularly fine book, 
{\em Prime numbers, a computational perspective}~\cite{CP}, 
whose well-thumbed first edition has been a source of delight to me for 15 years. 
His work on {\em Integrals of the Ising class}~\cite{BBC} led him to high-precision evaluation of Bessel moments.
In Richard's memory, I offer a striking conjecture that generalizes~(\ref{sr2}), to an astounding
degree, and has resulted in the discovery of a 204433-digit probable prime divisor of a Bessel moment.

{\bf Conjecture 5.} For positive integer $n$, let
\begin{equation}
A(n):=\left(\frac{2}{\pi}\right)^4
\int_0^\infty\left(\pi^2I_0^2(t) - K_0^2(t)\right) I_0(t)K_0^5(t)\,(2t)^{2n-1} dt.
\label{An}\end{equation}
Then $A(n)$ is a non-negative integer and for $1<n<x$  the number of cases for which
precisely one prime $p\ge n/2$ divides $A(n)$ is asymptotic to
\begin{equation}
N(x):=\frac{\exp(\gamma)[\log(x/2)]^2}{4+8\log(2)}.
\label{Nx}\end{equation}

{\bf Remarks.} From numerical integration, at high precision, I found that, very probably,
$A(n)$ gives a sequence of integers, beginning with\\
{\tt 0, 1, 2, 15, 302, 12559, 900288, 98986140, 15459635718}\\
for $n=1$ to 9, with $A(1)=0$ corresponding to sum rule~(\ref{sr2}).
Then I used the recursion
\begin{equation}
\sum_{k=0}^4(-1)^k(k+1)P_k(2n+k)A(n+k)=0
\label{rec}\end{equation}
with polynomial coefficients  $P_0(x)=x^9$,
$P_1(x)=15x^7+63x^5+45x^3+5x,$
$P_2(x)=91x^5+365x^3+188x,$
$P_3(x)=205x^3+371x$
and $P_4(x)=\frac{576}{5}x$,
determined by the differential equation satisfied by both $I_0(t)$ and $K_0(t)$,
discovering that the (presumed) integers $A(n)$ are rich in odd prime factors $p<n/2$. 
For example,
\begin{eqnarray}
A(33)&=&2\times3^{23}\times5^{11}\times7^6\times11^2\times13^2\times p_{56}\label{A33}\\
A(36)&=&3^{25}\times5^{13}\times7^8\times11^4\times13^2\times17^2\times p_{59}\label{A36}\\
A(49)&=&2^3\times3^{39}\times5^{16}\times7^{10}\times11^6\times13^4\times17^2\times
19^3\times23^2\times p_{86}\label{A49}\quad{}\quad{}
\end{eqnarray}
where $p_{56}$ is the 56-digit prime\\
{\tt 57992474894877287439798522082574263282518819530344295461.}\\
Similarly, $p_{59}$ and $p_{86}$ are primes with 59 and 86 decimal digits.

Now consider the sequence of integers $n>1$ for which $A(n)$ has precisely
one prime divisor $p\ge n/2$. It begins with {\tt 3, 5, 7, 9, 11, 33, 36, 49}\phantom{,} and
contains 5 more integers with $n\leq2000$, namely
{\tt 453, 727, 1560, 1569, 1627,} for which proofs of primality
were relatively easy.  There are 8 cases with $2000<n<10^5$ for which $A(n)$ has only 
one probable prime divisor $p\ge n/2$. They are 
{\tt  5078, 6605, 17663, 27281, 29298, 29708, 39509, 98653.}
Here I relied on strong Lucas and Fermat tests of probable prime divisors
with up to 204433 decimal digits.

{\bf Heuristics.} Asymptotic expansion of the Bessel functions in~(\ref{An}) gives
\begin{equation}
A(n)=\frac{8(2n-5)!}{\pi}\left(1
-\frac{1}{4n}
-\frac{3}{32n^2}
+\frac{141}{128n^3}
+\frac{14019}{2048n^4}
+O\left(1/n^5\right)\right)
\label{asy}\end{equation}
which grows factorially, with $A(60000)$ a 557365-digit integer.

For $n>1$, let $A(n)=B(n)C(n)$, where no prime $p\ge n/2$
divides the integer $B(n)$ and no prime $p<n/2$ divides
the integer $C(n)>0$, which grows more slowly, with
$C(60000)$ a 124405-digit integer. Empirically, it seems that
\begin{equation}\log(C(n))=n\log(K)+O(\sqrt{n})
\label{emp}\end{equation}
with $K\approx118$ and a fluctuating term suppressed by a square root.

Here I shall argue that
\begin{equation}K=(4e)^2=118.22489758\ldots\label{Kis}\end{equation}
For odd prime $p<n/2$ let $k$ be the largest integer such that $p^{2k}|B(n)$.
Then experiment reveals that $k\ge\lceil n/p\rceil-2$ and that
this lower bound is rarely exceeded for $p>\sqrt{n}$. For
$\sqrt{n}\ge p>2$, the empirical estimate $k=(n+O(\sqrt{n}))/(p-1)$ is adequate
for the present purpose. Combining these observations,
I obtain~(\ref{Kis}) using the following sums over primes:
\begin{eqnarray}
\sum_{p<x}\frac{\log(p)}{p-1}
&=&\log(x)-\gamma+o(1)\label{gamm}\\
\sum_{\sqrt{x}<p<x}\left(\left\lceil\frac{x}{p}\right\rceil-\frac{x}{p}\right)\frac{\log(p)}{x}
&=&\gamma+o(1).\label{gamp}
\end{eqnarray}
Mertens' third theorem then
gives the probability of $C(n)$ being prime, heuristically, as
asymptotic to $\exp(\gamma)\log(n/2)/\log(K^n)$
and the number of prime values of $C(n)$ with $1<n<x$ as asymptotic to $N(x)$,
as defined in (\ref{Nx}), which gives $N(60000)\approx19.8$. This is comfortingly
close to the number 20 of probable primes that I discovered with $n\in[2,60000]$.

Work by Armin Straub confirmed my findings up to $n=17663$. Then
Paul Zimmermann and Bruno Salvy confirmed them up to $n=39509$.
I encountered an unexpectedly
long gap until the next case, at $n=98653$, where the  204433-digit
probable prime $C(98653)$
and was found on 15 November 2015, after considerable work with~{\tt OpenPFGW}.
Neil Sloane has recorded $98653$ as entry 21 in sequence A265079
of {\em The on-line encyclopedia of integer sequences}~\cite{Soeis}.

\newpage

\subsection{Evaluation of an L-series of weight 6 outside its critical strip}%7.8

The rationality of the moments $A(n)$, asserted in Conjecture~5,
was a source of joy, since it implies rich rational substructure in matrices
of moments of 8 Bessel functions. We had hoped for a $3\times3$ matrix,
involving 7-loop Feynman integrals, with a determinant that might yield
$L_8(7)$, outside the critical strip, where a Mahler measure had failed to do the job.
Thanks to the conjectures that $A(1)$, $A(2)$ and $A(3)$ evaluate to
the integers 0, 1 and 2, respectively, we found something even better.

{\bf Conjecture 6.} The determinant of the $2\times2$ matrix
\begin{equation}
{\cal M}_2:=\int_0^\infty K_0^6(t)\left[\begin{array}{cc}
K_0^2(t)&t^2(1-2t^2)K_0^2(t)\\
I_0^2(t)&t^2(1-2t^2)I_0^2(t)\end{array}\right]t\,dt\,
\label{con6m}\end{equation}
with 8-Bessel moments up to 7 loops,
evaluates the L-series~(\ref{L8}) for the weight-6 modular form~(\ref{f8}),
outside its critical strip, as follows:
\begin{equation}
L_8(7)=\frac{128\pi^2}{6075}\det{\cal M}_2.
\label{con6d}\end{equation}

\subsection{Further evaluations of determinants of Feynman integrals}%7.9

Now consider the $3\times3$ matrix
\begin{equation} 
N_3:=\int_0^\infty I_0(t)K_0^5(t)\left[\begin{array}{rrr}
K_0^2(t)&t^2K_0^2(t)&t^4K_0^2(t)\\
I_0(t)K_0(t)&t^2I_0(t)K_0(t)&t^4I_0(t)K_0(t)\\
I_0^2(t)&t^2I_0^2(t)&t^4I_0^2(t)\end{array}\right]t\,dt
\label{N3}\end{equation}
obtained by adding an extra $K_0(t)$ to the integrand in~(\ref{M3}). 
The elements of its first column are evaluated
by $S_{8,6}$, in~(\ref{S86}), by $S_{8,5}$, in~(\ref{S85}), and by $S_{8,4}$, in~(\ref{S84}).
The elements in its first row are related to those in its third row, by Conjecture~5.
Thus we have 5 relations constraining the 9 elements.  There is a 6th empirical constraint:
\begin{equation}
\det N_3\;\stackrel{?}{=}\;\frac{5}{3}\frac{\pi^8}{2^{19}}
\label{detn3}\end{equation}
which has likewise been checked at high precision
and is the tip of another empirical iceberg.

Let $N_k$ be the $k\times k$
matrix with elements
\begin{equation}
(N_k)_{a,b}:=\int_0^\infty[I_0(t)]^a[K_0(t)]^{2k+2-a}t^{2b-1}dt,
\label{Nk}\end{equation}
which are moments of $2k+2$ Bessel functions.  For integer $m>0$, let
\begin{equation}
D_m:=\frac{2\pi^{m^2/2}}{\Gamma(m/2)}\prod_{j=1}^m\frac{(2j-1)^{m-j}}{(2j)^j},
\label{Dm}\end{equation}
which is a rational multiple of an integer power of $\pi^2$.

{\bf Conjecture 7.} For every integer $k>0$, the determinant of $N_k$ is $D_{k+1}$. 

{\bf Examples.} The constant $D_1=1$ is not a subject of the conjecture; one may think of it
as referring to the empty matrix. $D_2=(\pi/4)^2$ correctly
evaluates the two-loop integral $\int_0^\infty I_0(t)K_0^3(t)\,t\,dt=S_{4,2}/4$ defined 
by~(\ref{Nk}) at $a=b=k=1$. $D_3=\frac{5}{32}\zeta(4)$ gives conjecture~(\ref{con6i}),
for 6 Bessel functions, and $D_4=\frac53\pi^8/2^{19}$ 
gives conjecture~(\ref{detn3}), for 8 Bessel functions.
I have checked Conjecture~7 up to $k=15$, where a matrix of Feynman
integrals with up to 30 loops is predicted to have determinant
\begin{equation}
D_{16}=17^7\left(\frac{19}{7}\right)^6\left(\frac{23}{65}\right)^4
\frac{7}{3}\,\frac{29}{11}\,\frac{\pi^{128}}{2^{291}}
\label{det16}\end{equation}
in agreement with 225 numerical quadratures at 500-digit precision.

\section{Harder problems at $n>8$}

The bad news is that, despite strong effort, I have not yet evaluated
an L-series for $n>8$ Bessel functions in terms of Feynman integrals.
Yet I feel that it may still be possible. My intuition is based on the
following remarkably tight structure, imposed by Conjecture~2 on the case $n=9$.

For the 52 primes in the set ${\cal S}:=\{2<p\leq631,\left(\frac{p}{105}\right)=1\}$, I found that 
\begin{eqnarray}
\frac{p\,c_9(p)}{4}&=&\left\{\begin{array}{rl}
b_p^2-945a_p^2&\mbox{if }p\equiv1\mbox{ mod }3\\
21b_p^2-45a_p^2&\mbox{if }p\equiv2\mbox{ mod }3\end{array}\right.\label{c9m}\\
\frac{4p^6+c_9^2(p)-c_9(p^2)}{16p^2}&=&\left\{\begin{array}{rl}
b_p^2+945a_p^2&\mbox{if }p\equiv1\mbox{ mod }3\\
21b_p^2+45a_p^2&\mbox{if }p\equiv2\mbox{ mod }3\end{array}\right.\label{c9p}
\end{eqnarray}
where $(a_p,b_p)$ is a pair of positive integers, determined by $c_9(p)$ and $c_9(p^2)$.
For each of these 52 primes, I define an integer measure, $\alpha_p$, of how close $c_9(p)$ comes 
to determining $c_9(p^2)$, as follows. 

For  $p\in{\cal S}$, let $\alpha_p$ be the number of positive integers $m<a_p$
such that $p\,c_9(p)/4+945m^2$ is a square, for $p\equiv1$ mod 3, or 
$(p\,c_9(p)/4+45m^2)/21$ is a square, for $p\equiv2$ mod 3.
If $\alpha_p=0$, I say that $c_9(p)$ determines $c_9(p^2)$, since we may
take $a_p$ as the smallest possible positive integer consistent with~(\ref{c9m}),
which then tells us the positive integer $b_p$ and hence $c_9(p^2)$, from~(\ref{c9p}). 
If $\alpha_p$ is positive, I say that $c_9(p)$ fails to determine $c_9(p^2)$,
via~(\ref{c9m},\ref{c9p}), and take the value of $\alpha_p$ as a measure
of how far it falls short. In either case, $\alpha_p$ and $c_9(p)$
determine $c_9(p^2)$. 

Here are the 52 values of $\alpha_p\leq4$,
\begin{equation}
{\tt 0000100131100001000002040000010000101020200000000111}
\label{s9}\end{equation}
concatenated as a string, with increasing $p$, in order to save space. 
To obtain these, I used
Algorithm~2, with $k=1$, for each of the 52 primes $p\in{\cal S}$, to determine
$c_9(p)$, which was quickly done. At $k=2$
\begin{equation}
2\sum_{p\in{\cal S}} (p^2-1)^2=3019245508224>3\times10^{12}
\label{big}\end{equation}
accesses to tables of traces of Frobenius were made
to determine $c_9(p^2)$, which took far longer, by a factor of more than $2\times10^5$.
Yet the results from all that long work on $c_9(p^2)$ are encoded in one line, 
by the 52 integers $\alpha_p\leq4$ listed in~(\ref{s9}),
of which 36 vanish, showing that, by my definition, $c_9(p)$ determines $c_9(p^2)$
{\em more often than not}, for $p\leq631$. It was the astounding data compression in~(\ref{s9})
that led me to Conjectures~1,~2 and~3. I remark that there is no evidence that this
phenomenon is limited to small primes. On the contrary, the average
value of $\alpha_p$ for the 26 smaller primes in ${\cal S}$ is about $0.54$,
while for the 26 larger primes it is about $0.38$.

{\bf Example 1.}
At $p=577$, we have $c_9(577)/4=-16930160$ and $\alpha_{577}=0$.
The {\em smallest} case with $-577\times16930160+945m^2$  a square is $m=3216$.\\
Then $(a_{577},b_{577})=(3216,2260)$ give $c_9(577^2)=100104812100156676$.

{\bf Example 2.}
At $p=617$, $c_9(617)/4=-5181267$ and  $\alpha_{617}=1$ determine
$m=11872$ as the {\em second} case with $(-617\times5181267+45m^2)/21$
a square. Then $(a_{617},b_{617})=(11872,12239)$ give
$c_9(617^2)=-92449542374608124$.

These show how a single line, in~(\ref{s9}), spares any future worker $99.9995\%$  of the effort that I have
expended at $n=9$. The functional equation~(\ref{Msym}) is very fine, but still lacking
in the structure implied by my Conjectures~1, 2 and~3, tested for $n<20$.

I conclude with questions~\cite{Br,K} posed in the introduction.
\begin{enumerate}
\item {\em What proverb more common, what proverb more true, than that 
after pride comes a fall?} 
\item {\em Ah, but a man's reach should exceed his grasp, or what's a heaven for?}
\end{enumerate}

{\bf Acknowledgements.} Many people have advised and encouraged me. 
Here I particularly thank
Spencer Bloch, 
Francis Brown, 
Freeman Dyson, 
Dirk Kreimer, 
Stefano Laporta, 
Anton Mellit, 
Oliver Schnetz,
Neil Sloane and 
Don Zagier, 
whose own efforts continue to inspire and inform mine.}

\end{document}